\documentclass[12pt,preprint]{aastex}

\newcommand{\ngc}{NGC~7469}
\newcommand{\kms}{km~s$^{-1}$}

\newcommand{\lya}{Ly$\alpha$}
\newcommand{\lyb}{Ly~$\beta$}

\newcommand{\fuse}{{\it FUSE}}
\newcommand{\stis}{STIS}
\newcommand{\chand}{{\it Chandra}}

\begin{document}

\title{Intrinsic Absorption in the Spectrum of NGC~7469:
Simultaneous {\it Chandra},
\fuse\, and \stis\ Observations}

\author{Jennifer E.\ Scott\altaffilmark{1}, Gerard A.\ Kriss\altaffilmark{2,3},
Julia C.\ Lee\altaffilmark{4,5},
Jessica Kim Quijano\altaffilmark{2},
Michael Brotherton\altaffilmark{6},
Claude R.\ Canizares\altaffilmark{7},
Richard F.\ Green\altaffilmark{8}, John Hutchings\altaffilmark{9},
Mary Elizabeth Kaiser\altaffilmark{3},
Herman Marshall\altaffilmark{7}, William Oegerle\altaffilmark{10},
Patrick Ogle\altaffilmark{11}, 
\& Wei Zheng\altaffilmark{3}}

\altaffiltext{1}{
Observational Cosmology Laboratory,
National Aeronautics and Space Administration,
Goddard Space Flight Center,
Greenbelt, MD 20771  USA;
jscott@stis.gsfc.nasa.gov}

\altaffiltext{2}{Space Telescope Science Institute, 3700 San Martin Drive,
Baltimore, MD  21218 USA; [gak,jkim]@stsci.edu}

\altaffiltext{3}{Center for Astrophysical Sciences, Department of 
Physics and Astronomy,
The Johns Hopkins University, Baltimore, MD 21218 USA; 
[kaiser,zheng]@pha.jhu.edu}

\altaffiltext{4}{Harvard-Smithsonian Center for Astrophysics,
60 Garden Street, MS-4,
Cambridge, MA 02138  USA;
jclee@cfa.harvard.edu}

\altaffiltext{5}{Chandra Fellow}

\altaffiltext{6}{Department of Physics and Astronomy, University of Wyoming,
Laramie, WY, 82071  USA; mbrother@uwyo.edu}

\altaffiltext{7}{Department of Physics and Center for Space Research,
Massachusetts Institute of Technology, 77 Massachusetts Avenue, NE80, Cambridge,
 MA 02139  USA;
[crc,hermanm]@space.mit.edu}

\altaffiltext{8}{Kitt Peak National Observatory, National Optical 
Astronomy Observatories, 
P.O. Box 26732, 950 North Cherry Avenue, Tucson, AZ 85726  
USA; rgreen@noao.edu}

\altaffiltext{9}{Herzberg Institute of Astrophysics, National
Research Council of Canada, Victoria, BC V9E 2E7, Canada; 
john.hutchings@hia.nrc.ca}

\altaffiltext{10}{ExoPlanets and Stellar Astrophysics Laboratory,
National Aeronautics and Space Administration,
Goddard Space Flight Center,
Greenbelt, MD 20771  USA;
oegerle@uvo.gsfc.nasa.gov}

\altaffiltext{11}{
Mail Code 238-332,
Jet Propulsion Lab,
4800 Oak Grove Drive,
Pasadena, CA 91109  USA;
pmo@sgra.jpl.nasa.gov}

\begin{abstract}

We present simultaneous X-ray, far-ultraviolet, and near-ultraviolet spectra
of the Seyfert 1 galaxy NGC~7469 obtained with the 
{\it Chandra X-Ray Observatory}, the
{\it Far Ultraviolet Spectroscopic Explorer}, and the Space Telescope Imaging
Spectrograph on the {\it Hubble Space Telescope}. Previous non-simultaneous
observations of this galaxy found two distinct UV absorption components,
at $-560$ and $-1900$ \kms, with the former
as the likely counterpart of the X-ray absorber.
We confirm these two absorption components in our new UV observations,
in which we detect prominent \ion{O}{6}, Ly$\alpha$, \ion{N}{5},
and \ion{C}{4} absorption.
In our Chandra spectrum we detect \ion{O}{8} emission, 
but no significant \ion{O}{8} or
\ion{O}{7} absorption. We also detect a prominent 
Fe K$\alpha$ emission line in the
Chandra spectrum, as well as absorption due to hydrogen-like and helium-like
neon, magnesium, and silicon at velocities consistent with the $-560$ \kms\
UV absorber. The \fuse\ and \stis\ data reveal that the \ion{H}{1} and \ion{C}{4}
column densities in 
this UV- and X-ray- absorbing component have increased over time, as the UV
continuum flux decreased. We use measured \ion{H}{1}, \ion{N}{5}, \ion{C}{4}, 
and \ion{O}{6} column
densities to model the photoionization state of both absorbers
self-consistently. 
We confirm the general physical picture of the outflow in
which the low velocity component is a highly ionized,
high density absorber with a total column density of
10$^{20}$ cm$^{-2}$, 
located near the broad emission line region, although due
to measurable columns of \ion{N}{5} and \ion{C}{4}, 
we assign it a somewhat smaller
ionization parameter than found previously, $U\sim 1$.
The high velocity UV
component is of lower density, $\log N=18.6$,
and likely resides farther from the central
engine as we find its ionization parameter to be $U= 0.08$.

\end{abstract}


\section{Introduction}

Absorption edges are visible in the X-ray spectra of
about one half of all low redshift AGN
(Reynolds 1997; George et al.\ 1998; Crenshaw, Kraemer, \& George 2003).
Ensemble studies of the warm absorber
population in Seyfert
galaxies (Blustin et al.\ 2005) indicate, from estimates of the radial
positions of the absorbers, that these
systems are likely predominately outflows from the dusty torus surrounding
the accretion disk in each object,
while those seen in PG quasars are more likely to arise in an accretion disk
wind.
All of these objects also show
high-ionization absorption lines in their UV spectra
\cite{cren1999}.  This suggests a connection between
the two phenomena, if not a complete identification of one with the
other.
The position of the absorbing complexes in velocity space, generally
blueshifted with respect to the systemic redshift of the parent AGNs, 
the variability of the absorption 
(Kriss et al.\ 1995; Shields \& Hamann 1997; Crenshaw \& Kraemer 1999;
Crenshaw et al.\ 2000;
Kraemer et al.\ 2001, 2002; Gabel et al.\ 2003a) 
and the non-unity continuum and/or broad emission line region (BELR) 
covering fractions
determined for many UV absorbers 
(Kriss et al.\ 2000b;
Gabel et al.\ 2003b;
Kraemer, Crenshaw, \& Gabel 2001;
Kraemer et al.\ 2001;
Kraemer et al.\ 2003;
Brotherton et al.\ 2002;
Crenshaw et al.\ 2003)
point to an origin for the gas that is intrinsic to the AGN.
AGN outflows are a potentially significant source of feedback
energy (Granato et al.\ 2004; Scannapieco \& Oh 2004), 
influencing the luminosity-temperature relation
for the intracluster medium of galaxy clusters by steepening
it on the scale of galaxy groups (Cavaliere et al.\ 2002), as well as
providing metals to 
and governing the formation of galaxies and
later generations of AGNs from the 
intergalactic medium 
(Adelberger et al.\ 2003).

Intrinsic absorption in many AGNs has been studied using high spectral
resolution UV and X-ray observations:
Mrk~279  (Scott et al.\ 2004; Gabel et al.\ 2005);
Mrk~509  (Kriss et al.\ 2000b; Kraemer et al.\ 2003; Yaqoob et al.\ 2003);
NGC~3516 (Kraemer et al.\ 2002);
NGC~3783 (Kraemer, Crenshaw, \& Gabel 2001; Kaspi et al.\ 2001;
 Blustin et al.\ 2002; Gabel et al.\ 2003a,b);
NGC~4051 (Collinge et al.\ 2001; Ogle et al.\ 2004);
NGC~4151 (Crenshaw et al.\ 2000; Kraemer et al.\ 2001);
NGC~5548 (Mathur et al.\ 1999; Crenshaw \& Kraemer 1999;
 Brotherton et al.\ 2002; Kaastra et al.\ 2000, 2002; 
Crenshaw et al.\ 2003; Steenbrugge et al.\ 2003).
However, with the exception of Mrk~279 and some data taken for the
long-term campaign on NGC~3783, 
these observations have
been performed at different times in the UV and X-rays. 
The high degree of variability in AGN continua
complicates self-consistent photoionization modeling 
of the intrinsic absorption
from these non-simultaneous data, preventing firm
conclusions about the nature of the relationship between UV and X-ray
absorbers.

The Seyfert 1 galaxy \ngc\ ($z=0.01639$, de Vaucouleurs et al.\ 1991)
has been studied extensively
in the optical, ultraviolet, and X-ray regimes (Wanders et al.\ 1997;
Collier et al.\ 1998),
and has been the subject of simultaneous
UV and X-ray variability studies (Kriss et al.\ 2000a; 
Nandra et al.\ 1998, 2000).
The intrinsic absorption in \ngc\ has been studied previously
in the X-ray with {\it XMM-Newton} (Blustin et al.\ 2003) and in the UV 
with \fuse\ (Kriss et al.\ 2003).
However those observations were separated by one year, making the
campaign presented here the first set of simultaneous, high-resolution
UV and X-ray spectral observations.
These authors found two primary components in the UV, with outflow
velocities of $-569$ and $-1898$~\kms.  The low velocity component is
identified with the high-ionization phase of the 
X-ray absorption, also responsible for the X-ray emission.
Its partial covering fraction suggests its position coincides with
that of the BELR.
The high velocity UV component has no associated X-ray absorber. Likewise,
the low-ionization phase of the 
X-ray absorption appears to have no UV counterpart.

Here, we present the simultaneous UV and X-ray observations and describe
the properties of the AGN outflow.  We present photoionization models
of the absorbing gas and draw conclusions about its geometry.

\section{X-ray: {\it Chandra}}
\subsection{Data}
\label{sec-cxobs}

We observed \ngc\ on 2002 December 12-13
for 150 ksec with the High Energy Transmission
Grating Spectrometer (HETGS) on the {\it Chandra X-Ray Observatory}.
The HETGS
\footnote{For further details on
{\it Chandra} instruments, see http://cxc.harvard.edu/proposer/POG.}
provides resolving power up to $\sim$1000
from 0.5~keV to 8.0~keV through the use of two gratings:
(1) the Medium Energy Grating (MEG)
which gives good spectral coverage from
0.5~keV to 5.0~keV (2.5-25~\AA);
and (2) the High Energy Grating (HEG), which
is optimized for the
0.8-8.0~keV (1.6-15~\AA) spectral region.

We extracted the spectra using the standard
CIAO v3.0.1 pipeline.
In order to maximize signal-to-noise,
we co-added the $\pm1$~order
MEG and HEG spectra and binned to $0.02$~\AA\
for analysis.  Coarser $0.06$~\AA\,  bins were used to determine the
continuum model. We used
ISIS~v.1.2.1\footnote{http://space.mit.edu/ASC/ISIS/}
(Houck \& Denicola 2000)
for spectral analysis. 
See Table~\ref{table-obs} for a summary of the \chand\
observation details.

\subsection{Analysis:
Continuum Fit and Spectral Features}
\label{sec-xray}

\begin{figure}
\epsscale{0.8}
\plotone{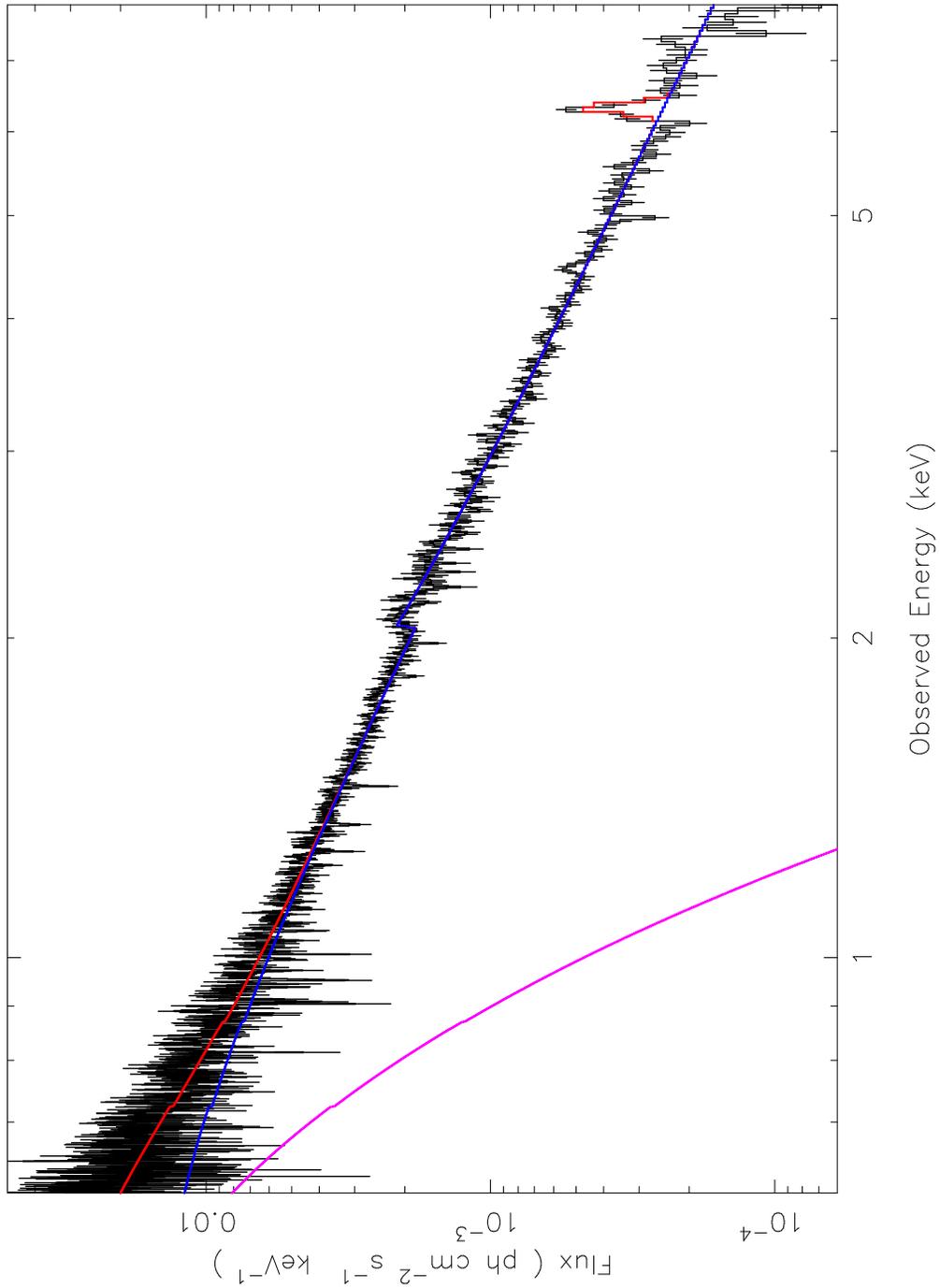}
\caption{{\it Chandra} HETG spectrum of \ngc, binned to
0.02~\AA, 
with best fit model including Fe K$\alpha$ emission (red).
Power law + iridium M-edge at 5.9 \AA\ (blue) 
and blackbody (magenta) components are shown separately.
\label{fig:chandra}}
\end{figure}

We fit the coarsely-binned MEG and HEG spectra 
in the 0.5-8.5~keV range
with a power-law plus blackbody
continuum modified by absorption from a Galactic hydrogen column
of $4.8\times 10^{20}$ cm$^{-2}$ (Dickey \& Lockman 1990).
We used the {\sc xspec} models {tbabs} (Wilms, Allen, \& 
M$^{\rm c}$Cray 2000) and 
{diskbb} (see, eg.\ Mitsuda et al.\ 1984; Makishima et al.\ 1986),
respectively to model 
the absorption and black-body components.
The photon index of the power-law spectrum 
is $\Gamma=1.7910\pm0.0009$, 
with a normalization of $A=(6.97\pm0.10) \times 10^{-3}$ 
photons cm$^{-2}$ s$^{-1}$ keV$^{-1}$
at 1 keV in the emitted frame, and the temperature of the
blackbody is $120\pm5$ eV, with overall
$\chi^2 / dof=1.12$.
As Blustin et al.\ (2003), 
we note that this continuum model is simply a convenient parametrization
that allows us to fit the emission and absorption features
in the \chand\ spectrum, 
it is not meant to be a physical model of the continuum.
The errors quoted are 90\% confidence limits.
The HETG spectrum, the continuum fit, and its components, including 
a Gaussian fit to the Fe K$\alpha$
line, are shown in Figure~\ref{fig:chandra}.
We also included additional multiplicative model components
to account for calibration effects.  We use:
(1) an ad hoc correction for ACIS pileup effects which also
reduces the effect of the iridium M-edge (Marshall et al.\ 2004a);
(2) a correction to bring the frontside-illuminated chip QEs
into agreement with the backside-illuminated chip quantum efficiencies
(Marshall et al.\ 2004a); and
(3) an ACIS contamination correction (Marshall et al.\ 2004b).

\begin{figure}
\epsscale{1.0}
\plotone{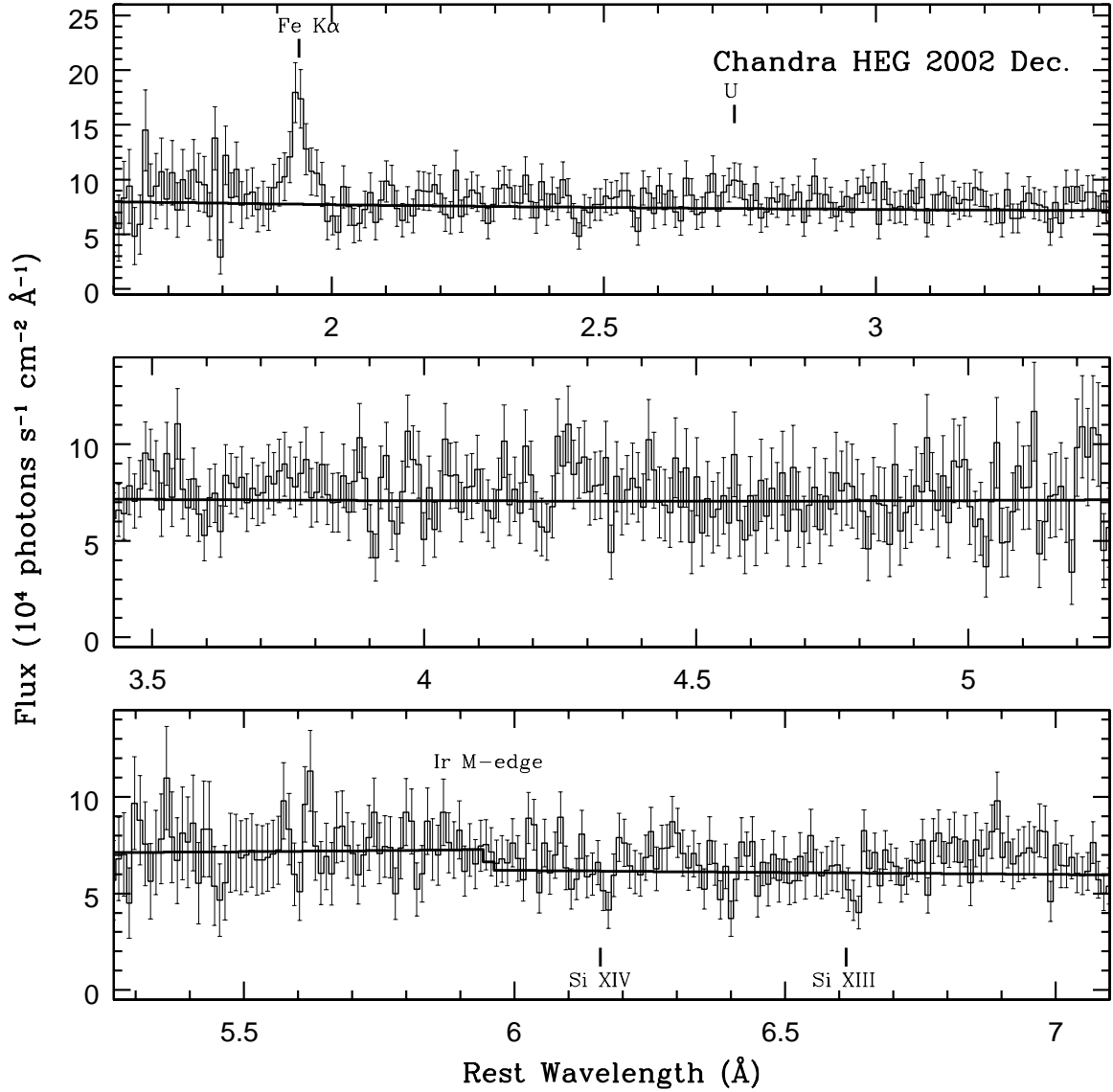}
\caption{HEG spectrum of NGC~7469 with flux errors
with absorption(emission) lines identified below(above) the
spectrum by tick marks with labels where ID known or $``$U" for
unidentified lines. 
The continuum fit
in Figure~\ref{fig:chandra} is shown by the solid line.  
\label{fig:heg1}}
\end{figure}

\begin{figure}
\epsscale{1.0}
\plotone{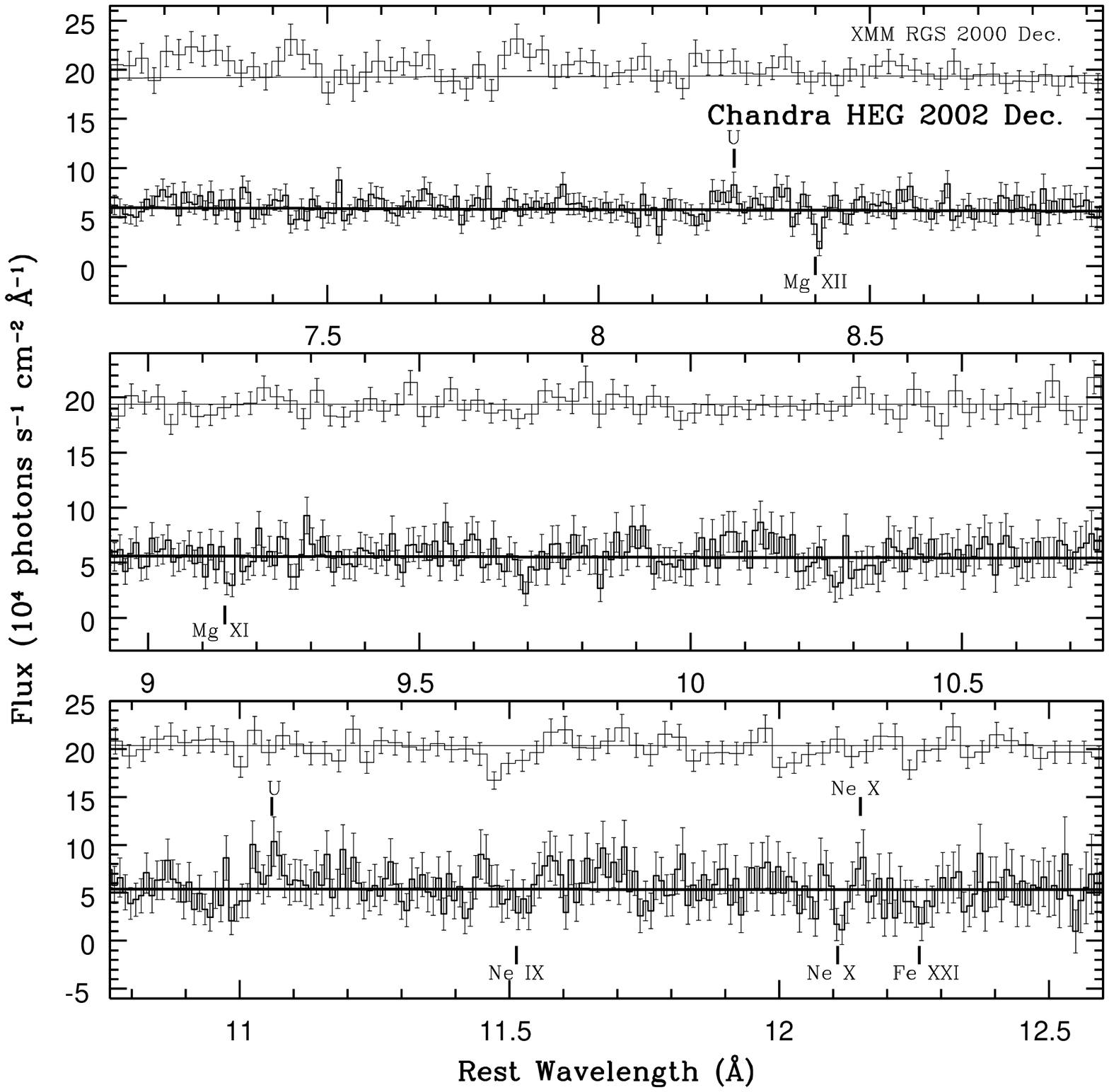}
\caption{HEG spectrum of NGC~7469 with flux errors
with absorption(emission) lines identified below(above) the
spectrum by tick marks with labels where ID known or
$``$U" for
unidentified lines. 
The continuum fit
in Figure~\ref{fig:chandra} is shown by the solid line.
Also shown is the {\it XMM}/RGS spectrum obtained in 2000 by
Blustin et al.\ (2003), offset by arbitrary factors for clarity.
\label{fig:heg2}}
\end{figure}

\begin{figure}
\epsscale{1.0}
\plotone{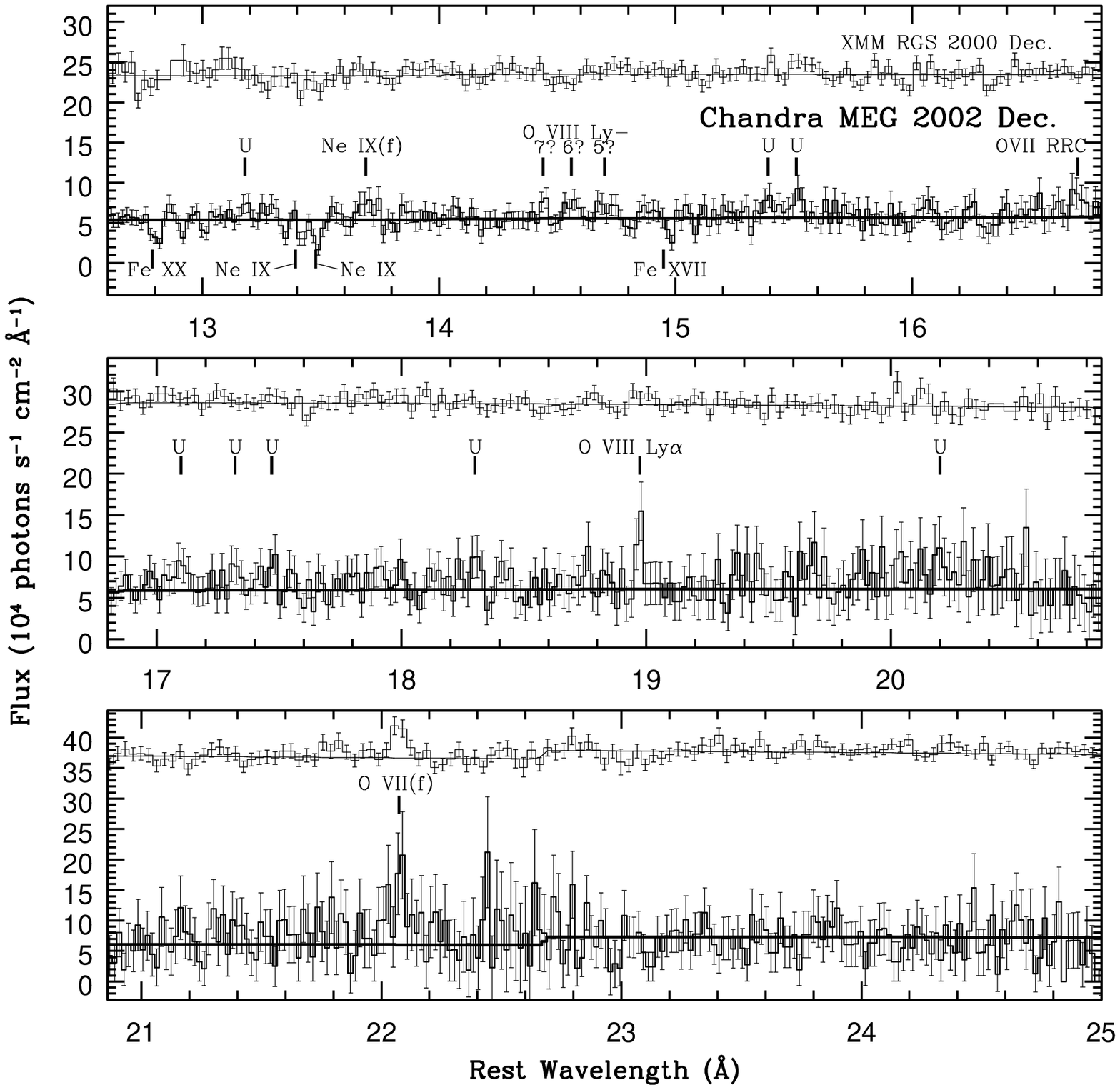}
\caption{MEG spectrum of NGC~7469  with flux errors
with absorption(emission) lines identified below(above) the
spectrum by tick marks with labels where ID known or $``$U" for
unidentified lines.
The continuum fit 
in Figure~\ref{fig:chandra} is shown by the solid line. 
Also shown is the {\it XMM}/RGS spectrum obtained in 2000 by
Blustin et al.\ (2003), offset by arbitrary factors for clarity.
\label{fig:meg1}}
\end{figure}

In regions of overlap,
we require a spectral feature to be identified in both the HEG and MEG
spectra in order 
to qualify as a bona fide absorption or emission line.
Because the resolution of the HEG spectrum is 2$\times$ that of the
MEG spectrum, we use the HEG spectrum alone to measure the centroid
equivalent width(flux) and FWHM of an absorption(emission)
line in the 1.6-12.6 \AA\ region.  At longer wavelengths,
the S/N in the HEG spectrum falls below $\sim$3 and we use the
MEG spectrum.  
We also require reasonable oscillator strengths for
quoted absorption features. For example, if we detect a Ly$\beta$ feature
from a particular species,
we require its Ly$\alpha$ transition to be detected as well.
The HEG and MEG spectra are shown in detail, with the absorption and
emission lines we identify and describe below, in 
Figures~\ref{fig:heg1}-\ref{fig:meg1}.  The continuum fit
shown in these figures is the global fit discussed above.
Because of the low count rate in the ACIS pixels, we
derive the flux errors per pixel 
from $\sigma_{N} \approx  1 + \sqrt{N + \frac{3}{4}}$
where $N$ is the count rate 
(Gehrels 1986), in order
to approximate Gaussian statistics. For comparison we
also show the {\it XMM}/RGS
spectrum from Blustin et al.\ (2003), in the regions of overlap
with the \chand/HETG data, $6-25$ \AA,
in Figures~\ref{fig:heg2} and \ref{fig:meg1}.

\begin{figure}
\epsscale{1.0}
\plotone{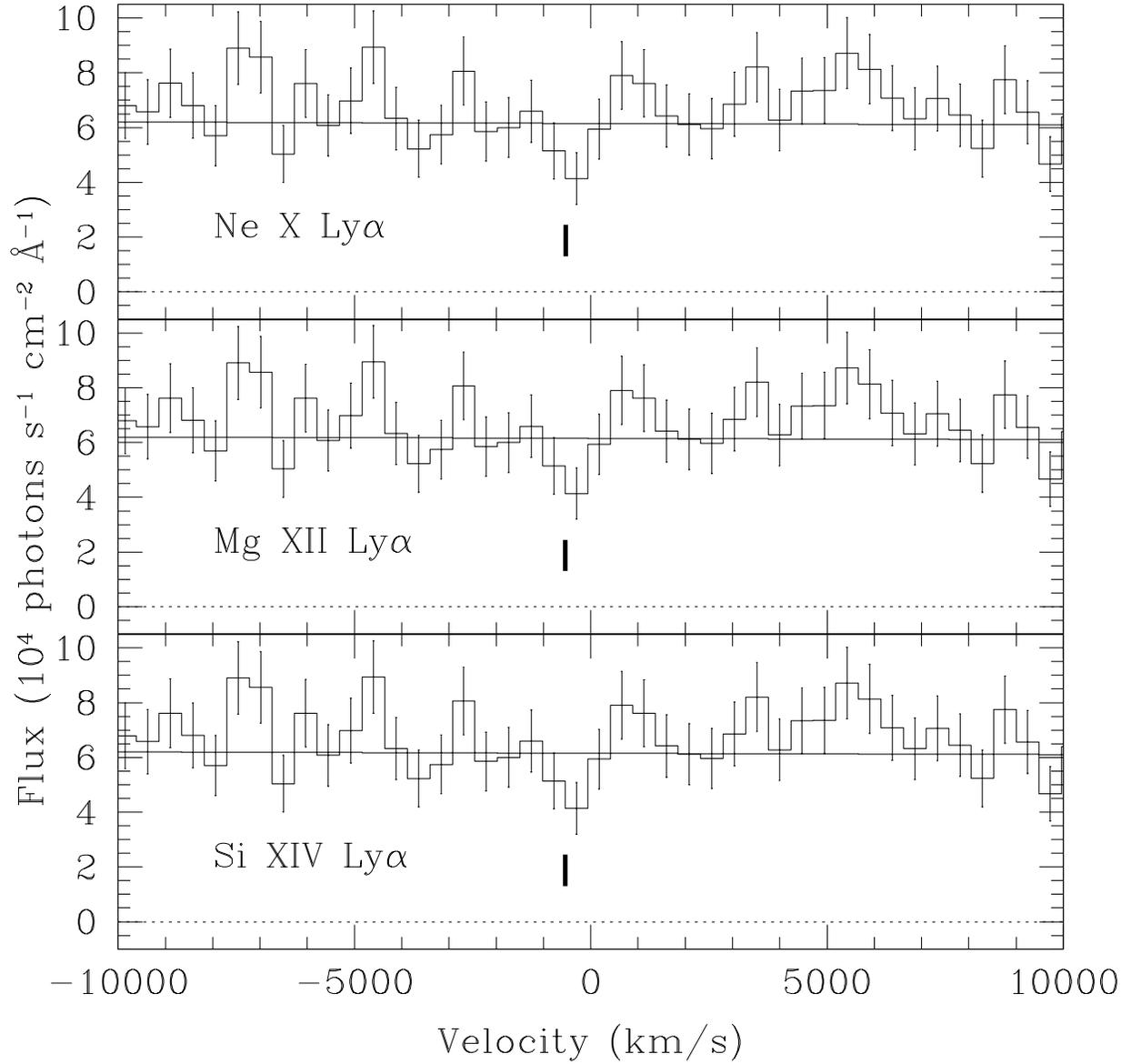}
\caption{Hydrogen-like lines in \chand/HETGS spectrum of \ngc, all from
the HEG channel.  The continuum level is
the global fit shown in Figure~\ref{fig:chandra}.
\label{fig-hlike}}
\end{figure}

\begin{figure}
\epsscale{1.0}
\plotone{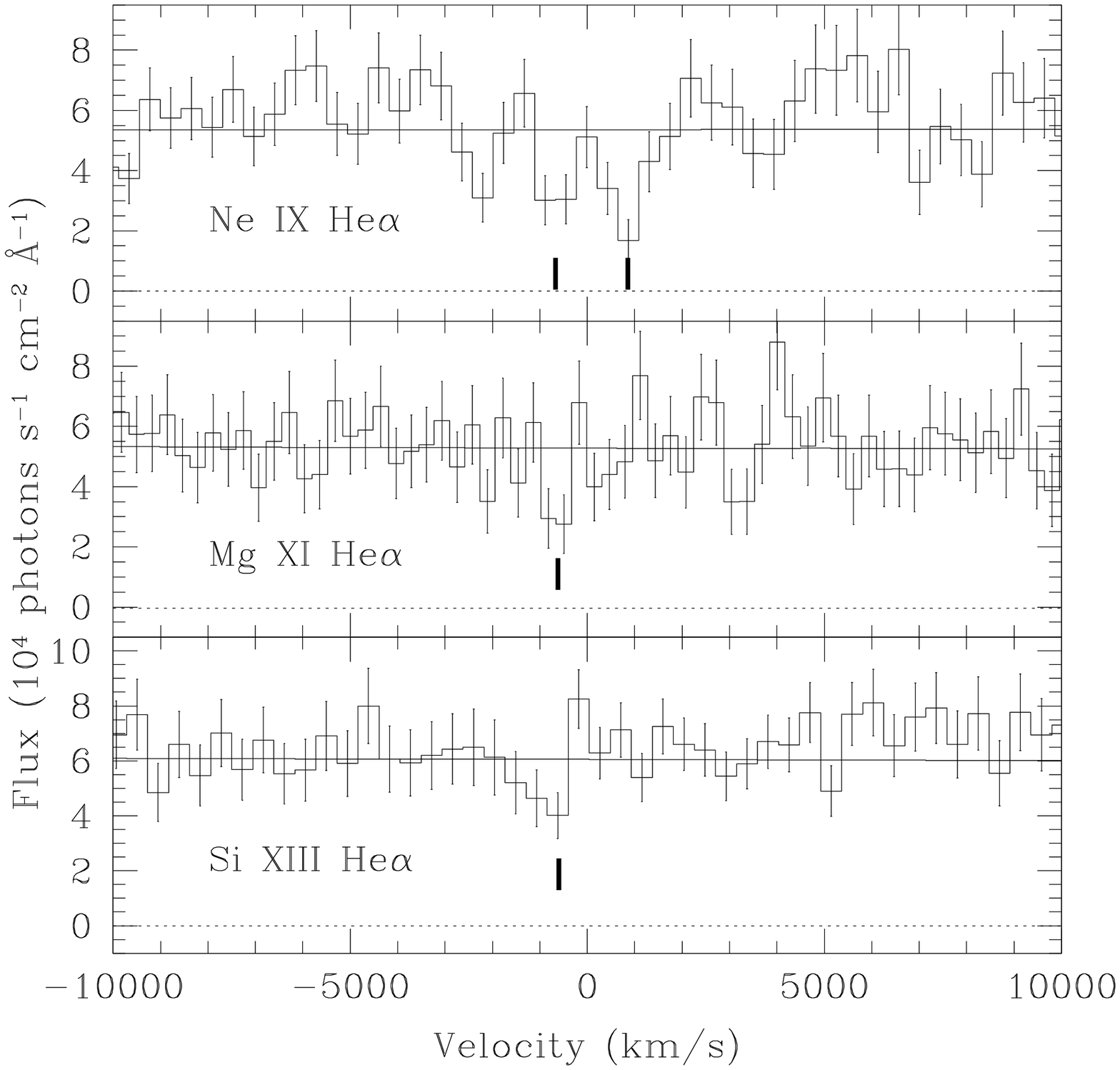}
\caption{Helium-like lines in \chand/HETGS spectrum of \ngc,
\ion{Ne}{9} at $v=-590$ and 860 \kms\ 
from MEG, \ion{Mg}{11} and \ion{Si}{13} from HEG
channel. The continuum level is
the global fit shown in Figure~\ref{fig:chandra}.
\label{fig-helike}}
\end{figure}

We find several absorption features in the \chand/HETG spectra,
and we fit Gaussians to the unresolved profiles to derive the equivalent
widths, the line centroids, and the velocity widths.
We find Ly$\alpha$ lines of the hydrogen-like ions
\ion{Ne}{10} and \ion{Mg}{12}, and a marginal \ion{Si}{14} feature 
(Figure~\ref{fig-hlike})
and the He$\alpha$ lines of 
\ion{Ne}{9}, \ion{Mg}{11}, and \ion{Si}{13} (Figure~\ref{fig-helike}).
We also detect \ion{Ne}{9} He$\beta$ with an outflow velocity and
line width consistent with the He$\alpha$ line. 
Of these features, Blustin et al.\ (2003) find only
\ion{Ne}{9} He$\beta$ in their {\it XMM}/RGS spectrum of
\ngc.  
However we note that the resolution of the RGS
is lower than that of both the HEG and the MEG, by factors
of $\sim$2.5 and $\sim$5 respectively.
Also,
the \ion{Si}{14} line lies at the high energy edge of the RGS spectral
range.
We also find three iron features,
\ion{Fe}{17} $\lambda$15.0, \ion{Fe}{21} $\lambda$12.3, 
and \ion{Fe}{20} $\lambda$12.8.
No \ion{Fe}{17} absorption is not found in the 
{\it XMM}/RGS data, but 
there is an ambiguous feature at the position of 
\ion{Fe}{21}. The \ion{Fe}{20} feature falls on a gap in the data.
Conversely, Blustin et al.\ (2003) report several 
absorption features that we do not find, or find
only marginal evidence for, in the \chand\ spectra:
\ion{O}{8} \lya\ and \lyb\, which we discuss further below, 
\ion{O}{7} 3-1 (r) (18.6 \AA), 
and \ion{Ne}{7} Be$\alpha$ (13.8 \AA).   The other lines
reported by these authors lie in regions where the HETGS
effective area is low, or  outside its spectral
range altogether.

The parameters of the absorption 
line fits are summarized in Table~\ref{table-cxo}, 
as well as those for the emission features discussed
below.  Along with the equivalent widths, line centroids, and
line widths, we tabulate the statistical significance of
each feature defined by the detection threshold.  This 
is defined as the equivalent width divided by the
the 1$\sigma$ equivalent width error, calculated as a function
of wavelength using the spectrum flux and error 
boxcar-smoothed over $\sim 2.5 \times$ the FWHM of the
line spread function (Bechtold et al.\ 2002).
We list all the absorption and emission
lines we detect at greater than $\sim2.5\sigma$ in Table~\ref{table-cxo},
including those for which we could make no identification.
The low confidence level we set leaves open the possibility that
some of the features in Table~\ref{table-cxo} are spurious.

We find a marginal narrow Ly$\alpha$ emission feature redward of the 
\ion{Ne}{10} absorption, shown in the top panel of
Figure~\ref{fig-hlike}.  The significance of both the
\ion{Ne}{10} absorption and emission lines in Table~\ref{table-cxo}
is low, 1.3$\sigma$, and they are not detected in the
{\it XMM}/RGS spectrum of Blustin et al.\ (2003).  However, 
we do detect them at 4.1$\sigma$ and 3.6$\sigma$
in the MEG spectrum.

\begin{figure}
\plotone{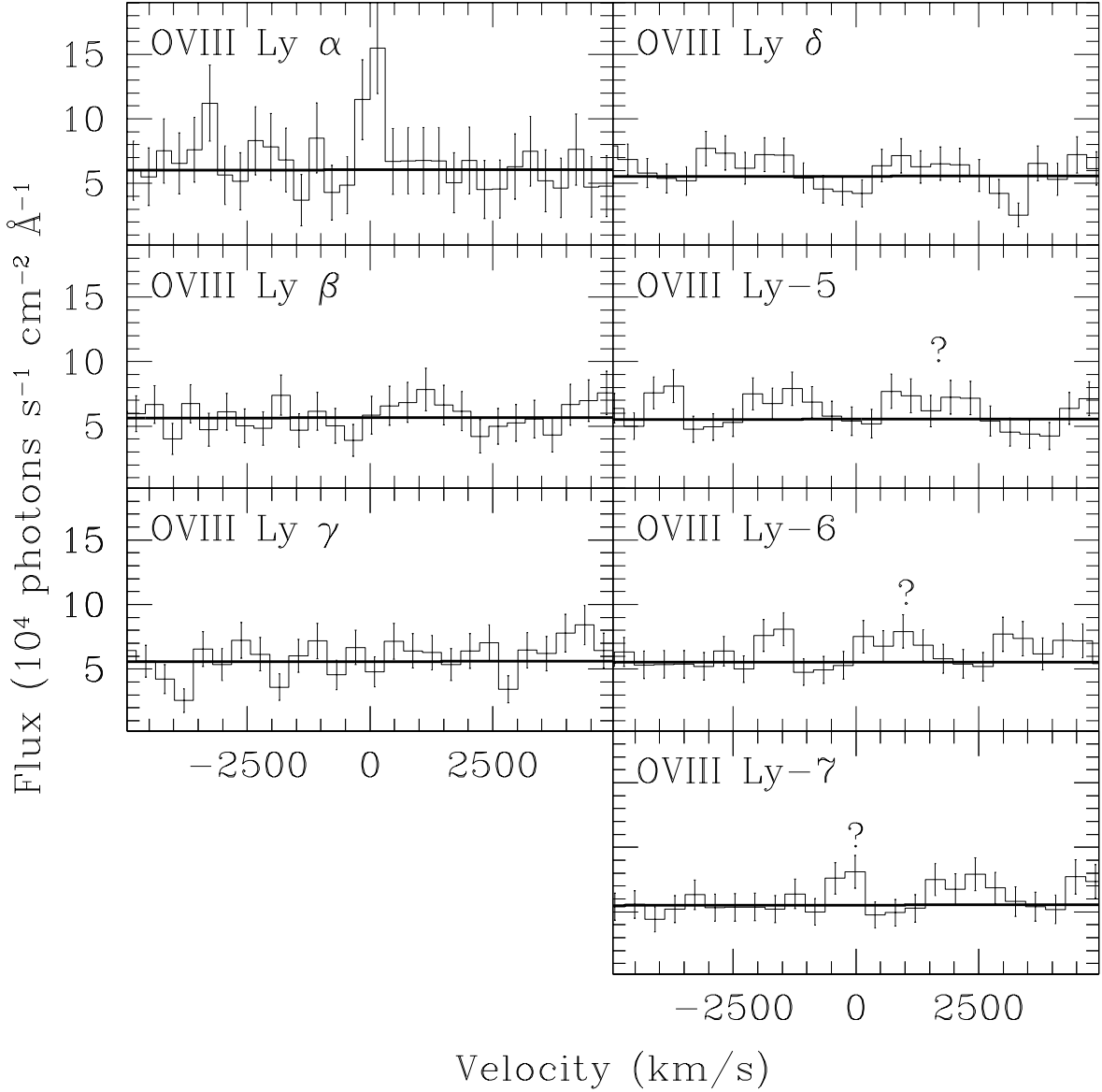}
\caption{Top left panel: 
\ion{O}{8} Ly$\alpha$ emission, detected at $3.5\sigma$,
and possible absorption
in the \chand\
MEG spectrum  of \ngc. Middle and bottom left panels and
top right panel show absence
of Ly$\beta$, Ly$\gamma$, and Ly$\delta$ features.  
Middle and bottom right panels show emission features
listed in Table~\ref{table-cxo}
tentatively identified as \ion{O}{8} Ly-5, Ly-6, and Ly-7.
The continuum level in all panels is
the global fit shown in Figure~\ref{fig:chandra}.
\label{fig-o8}}
\end{figure}

We see some evidence of \ion{O}{8} absorption blueward of, and
possibly filled in by
the Ly$\alpha$ emission line, as reported by Blustin et al.\ (2003),
but we do not find \ion{O}{8} Ly$\beta$ or Ly$\gamma$ absorption
in the {\it Chandra} spectrum (Figure~\ref{fig-o8}) as Blustin et al.\ (2003)
did.  We do not find
\ion{O}{8} Ly$\beta$, Ly$\gamma$, or Ly$\delta$ emission
corresponding to the prominent Ly$\alpha$ emission line, 
but we do find significant emission features near the
expected positions of the Ly-5, Ly-6, and Ly-7 lines.
However, because the implied velocities of these features
are not entirely consistent with the Ly$\alpha$ line,
these are marked as highly tentative 
identifications  in  Table~\ref{table-cxo}
and in Figure~\ref{fig-o8}.  These lines are not
seen in the {\it XMM}/RGS spectrum.

From the S/N in the spectrum at the
expected position of \ion{O}{8} Ly$\alpha$ absorption, we place a 3$\sigma$ 
upper limit of $\sim$30 m\AA\ on its
equivalent width, which
implies $N($\ion{O}{8}$)< 8.8 \times 10^{17}$~cm$^{-2}$ 
for $b \sim 100$ \kms.  
We find no \ion{O}{7} He$\alpha$ absorption, leading to 
an upper limit on the equivalent width of 80 m\AA, or
$N($\ion{O}{7}$)< 1.4 \times 10^{19}$~cm$^{-2}$, 
also for $b \sim 100$ \kms.
This choice of Doppler parameter derives from the identification of
this unresolved X-ray absorption with Component 1 in the UV spectra, which
we model with nine subcomponents with FWHM=61~\kms, as described in
Section~\ref{sec-intabs}.
If we instead assume $b=200$ \kms, the \ion{O}{8}
and \ion{O}{7} curves of growth give 
$N($\ion{O}{8}$)< 5.1 \times 10^{16}$~cm$^{-2}$ and
$N($\ion{O}{7}$)< 3.4 \times 10^{18}$~cm$^{-2}$,
while $b=500$ \kms gives
$N($\ion{O}{8}$)< 2.8 \times 10^{16}$~cm$^{-2}$ and
$N($\ion{O}{7}$)< 5.8 \times 10^{16}$~cm$^{-2}$. 
These dramatically lower \ion{O}{7} column densities do
not change the conclusions we draw from the photoionization
models in Section~\ref{sec-photo}.

\begin{figure}
\epsscale{1.0}
\plotone{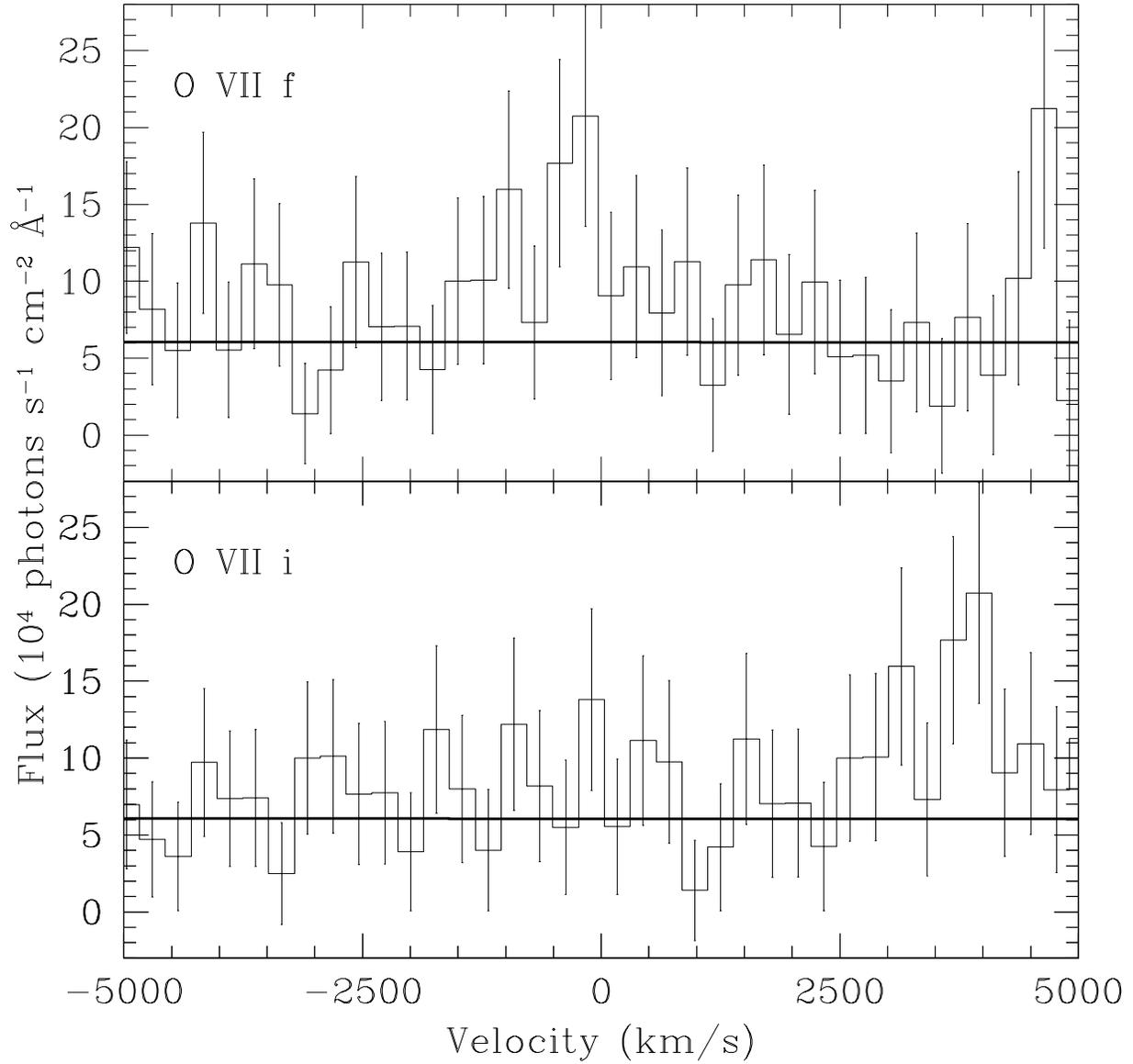}
\caption{\ion{O}{7} forbidden emission and
the expected position of corresponding intercombination emission
in the \chand\ MEG spectrum.
The continuum level is
the global fit shown in Figure~\ref{fig:chandra}.
\label{fig-o7}}
\end{figure}

An emission feature consistent with
\ion{O}{7} forbidden at $-200$~\kms\ 
is shown in the top panel of Figure~\ref{fig-o7}.
At $\sim$110 m\AA, it is stronger than reported by
Blustin et al.\ (2003).
The corresponding intercombination line, however, is highly uncertain
if it is present at all,
as shown in the bottom panel of the figure.
Blustin et al.\ (2003) report the forbidden to intercombination
ratio f/i~$\sim2$ for 
\ion{O}{7} in the {\it XMM} data from 2000. If that
ratio held for this observation, 
the S/N in the \chand\ spectrum is such that we would 
only expect to find the 
intercombination line at $\sim2.5\sigma$ confidence.
We find a significant emission feature near the 
expected position of the \ion{O}{7} radiative recombination
continuum, listed in Table~\ref{table-cxo}.
The emission feature in the top panel of Figure~\ref{fig-o7}
consistent with \ion{O}{7}(f) at $-980$ \kms\ is not statistically
significant. 
Finally, we identify a significant emission feature 
at 13.91 \AA\ observed with \ion{Ne}{9}(f) at
$-200$~\kms, although, again, we find no corresponding
resonance or intercombination lines.

A strong Fe K$\alpha$ emission line is present in the \chand\
spectrum. 
{\it BeppoSAX} observations (De Rosa et al.\ 2002) indicated
the presence of a narrow emission component in addition to the
broad emission from an accretion disk.
However, Blustin et al.\ (2003) found that the Fe K$\alpha$ emission line
in their {\it XMM} spectrum was best fitted by a single
narrow Gaussian.
Using the continuum defined above and 
fitting only the 1.5-2.5 \AA\ region of the
HEG spectrum, we find a single Gaussian with
flux $3.9 \pm 0.7 \times 10^{-5}$ photons cm$^{-2}$ s$^{-1}$,
line width $6310 \pm 1580$ \kms (FWHM), and rest-frame
energy $6.39\pm0.01$ keV  to be the best fit to this line
with $\chi^{2}/dof$= 62.5/57.  Adding a second, narrow 
Gaussian does not improve the fit.
These parameters are
consistent with the {\it XMM} results.
The width is better constrained by these
\chand\ data, however, and its value is consistent with the
UV emission lines we fit in the \stis\ spectra, particularly
the intermediate velocity width lines of Ly$\alpha$, \ion{C}{4},
and \ion{N}{5}.  These are discussed in the next section.
We note that Nandra et al.\ (2000) found a significant
correlation of the flux in the Fe K$\alpha$ line and the
2-10~keV continuum flux, indicating that at least some of the
flux arises within 1 light-day of the continuum source.

\section{Ultraviolet: {\it FUSE} and  STIS}

\subsection{Data}
We observed \ngc\ with \fuse\
using the $30\arcsec \times 30\arcsec$ low-resolution aperture.
See Table~\ref{table-obs} for the details of the observations.
We combined the two exposures taken on 2002 December 13
for a total exposure time of 6948 sec.
Because of visibility constraints imposed by the coordinated 
observation, there was insufficient time to align the SiC1, 
SiC2, and LiF2 channels for the \fuse\ exposures.
Our method for correcting for the
worm feature in the LiF1B segment uses the shape and flux level
of the target in the LiF2A channel, therefore we were not able to perform
the correction to derive a reliable spectrum over 
1093-1182 \AA. 

We used the standard \fuse\ calibration pipelines (CALFUSE v2.2.3, see
Sahnow et al.\ 2000), to extract the spectra and
to perform background subtraction and wavelength and flux calibrations.
We combined the spectra from each detector
and binned the spectra to 0.05~\AA\ bins, improving
the S/N while preserving the full
spectral resolution, $\sim$20~\kms.

To correct the wavelength scale of the \fuse\ spectra for zero-point offsets
due to placement of the target in the $30\arcsec$ aperture, we applied a 
0.0638 \AA\ shift to bring the wavelength scale into agreement with the heliocentric
spectrum observed by \fuse\ in 1999 described by Kriss et al.\ (2003). 
These authors established the zero point of the wavelength scale 
by comparing the positions of interstellar absorption lines such as \ion{Ar}{1},
\ion{Fe}{2}, \ion{O}{1}, and H$_{2}$ to the velocity of the 
Galactic 21 cm emission.
We estimate that the residual systematic errors in the FUSE wavelength scale
are on the order of 15 \kms, and that the flux scale is accurate to
$\sim$10\%.

Simultaneously with the \fuse\ observations,
we obtained observations of \ngc\ with the
\stis\ FUV MAMA over 5 orbits (13013 sec) through the
$0.2\arcsec \times 0.2\arcsec$ aperture
using the medium-resolution E140M echelle grating covering 1150-1730~\AA.
We also obtained 9 orbits (22810 sec) of \stis\ data on \ngc\ over two visits
on 2004 June 21-22 using the same set-up as above.
See Table~\ref{table-obs} for other details of the \stis\ observations.
We used CALSTIS v2.13b to process the spectra.
As described above, we measure the positions of lines from low-ionization
species, \ion{Si}{2}, \ion{N}{1}, \ion{Si}{3}, \ion{C}{2}, \ion{C}{2}$^{*}$, 
\ion{Si}{4}, \ion{Fe}{2},
and \ion{Al}{2}
in the \stis\ spectrum, and compare these to the mean heliocentric
velocity of the \ion{H}{1} 21-cm emission.  We find that no correction to the wavelength
scale derived by the CALSTIS pipeline is necessary.
Using a white dwarf spectrophotometric standard, 
we applied a small correction to the flux in the \stis\ spectra to correct for
residual echelle ripple structure and changes in the MAMA detector 
sensitivity not accounted for in the
current STIS pipeline. 
Gabel et al.\ (2005) 
will present a full description of this process. 
 
Finally, we discuss the origin of a depression in the \stis\ spectrum 
at 1418.8 \AA.  The expected position of redshifted \ion{Si}{4} at rest with
respect to the AGN is 1416.61 \AA, so this feature is inconsistent
with \ion{Si}{4} absorption due to intrinsic, outflowing gas.  Also,
this feature is shallow and broad, and qualitatively unlike other 
intrinsic features.  We attribute this feature to an instrumental effect,
namely a shadow on the MAMA detector.
The \stis\ FUV MAMA employs a field electrode, or repeller wire,
to direct any electrons emitted in directions away from the microchannel
plate detectors back into the channels.
The repeller wire shadow runs from $(x,y)=(1,994)$ to $(1048,1134)$.
The feature appears in order 104, which is
centered at $y=1030.7$ in pixel space, and runs from $y = 1038$ to $y= 1014$, 
covering
wavelengths 1414.3 to 1429.9 \AA.
The wire crosses the dispersion axis at an obtuse angle, thus the feature is
broad in the extracted spectrum.
Because the shadow appears in slightly different locations depending which
optical element is present in the spectrograph, it is not divided out
in the flat field, but simply flagged in the data quality array. 
A broad depression in the flux in STIS/E140M spectra
has been reported by others (Kraemer et al.\ 2001; Leitherer et al.\ 2001), and
a very similar feature is visible in an E140M spectrum of Mrk~279 (Gabel
et al.\ 2005).
Therefore, we conclude that this is not a feature intrinsic to \ngc.

\subsection{Analysis}
\subsubsection{Continuum and Emission Lines}
\label{sec-uvspec}

\begin{figure}
\epsscale{0.85}
\plotone{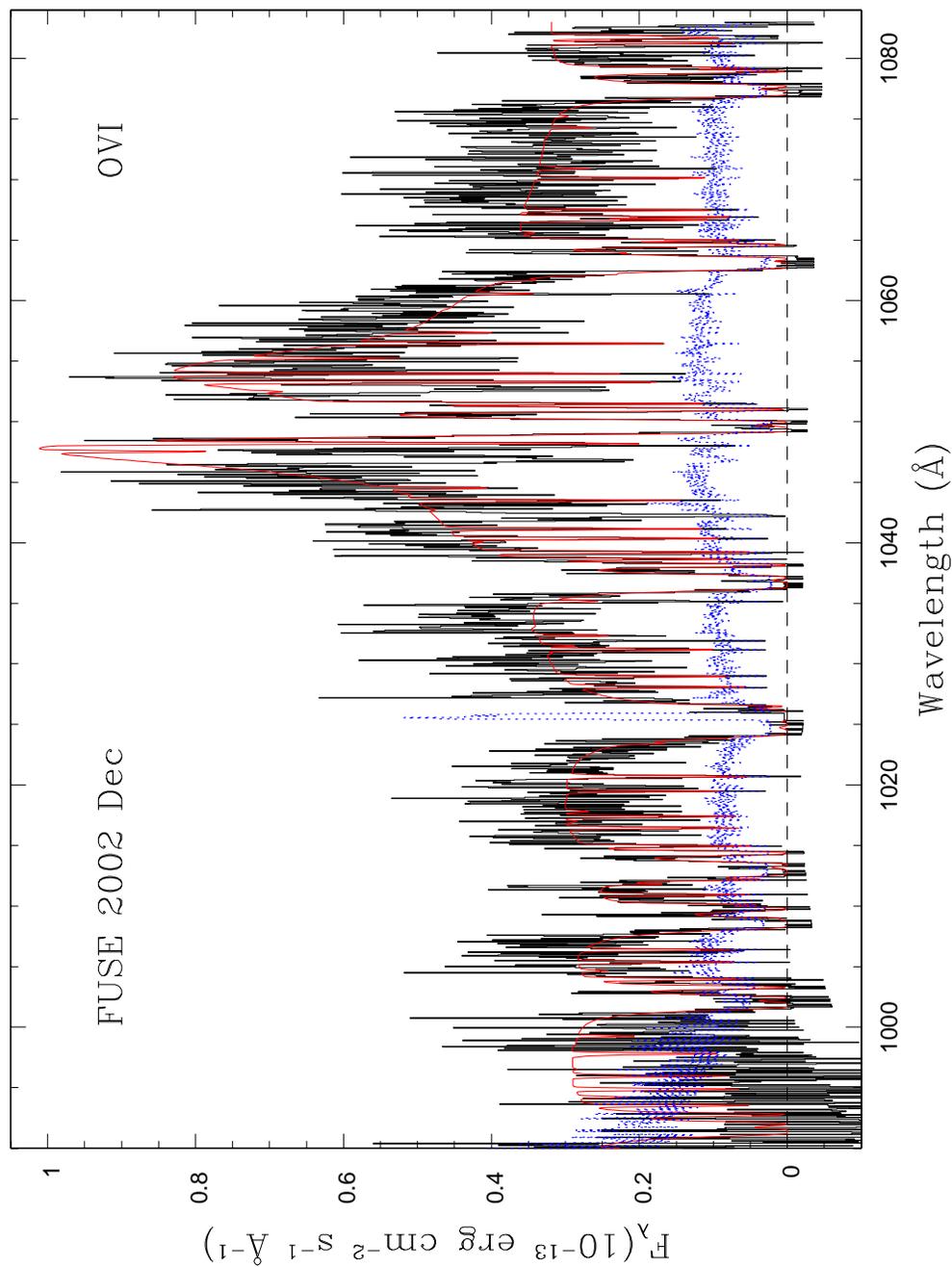}
\caption{The 2002 Dec \fuse\ spectrum of \ngc\ (black) with
errors shown in blue, and with
continuum, emission line, and ISM absorption fits (red).
\label{fig-fusespec}}
\end{figure}

\begin{figure}
\epsscale{0.85}
\plotone{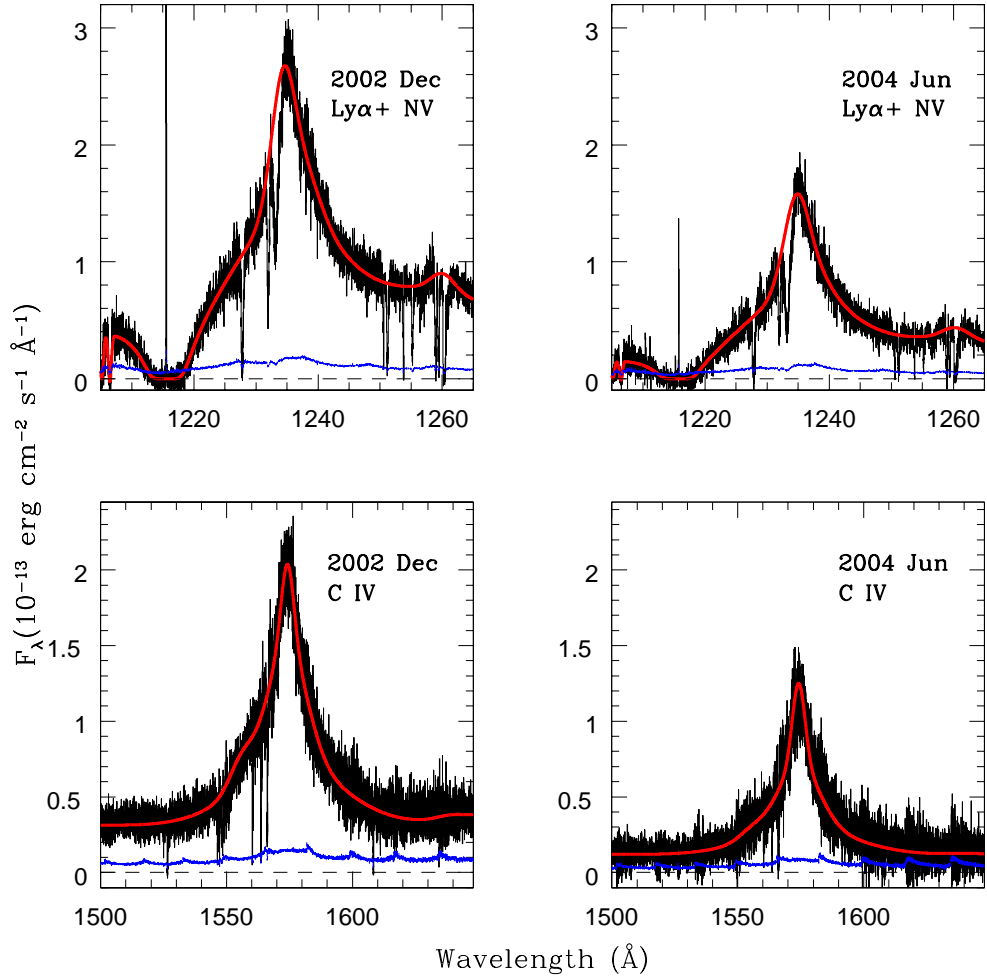}
\caption{The 2002 Dec and 2004 Jun \stis\ spectra of \ngc\ (black)
with errors shown in blue, and with
continuum and emission line fits (red).
\label{fig-stisspec}}
\end{figure}

We used the IRAF\footnote{IRAF is distributed by the National Optical
Astronomy Observatories,
which are operated by the Association of Universities for Research
in Astronomy, Inc., under cooperative agreement with the National
Science Foundation.}
task {\it specfit}  (Kriss 1994)
and spectral regions unaffected by absorption features
to fit the continuum and emission lines in the \fuse\ and \stis\ spectra
of \ngc.  This routine is an interactive spectral fitting tool
that uses $\chi^{2}$ minimization techniques to derive the best
fitting parameters for a user-specified model.
For the continuum, we fit
a power law
of the form $f_{\lambda} = f_{1000} \lambda^{-\alpha}$ to
the 2002 \fuse\ and \stis\ spectra simultaneously, where
the normalization, $f_{1000}$, is the flux at 1000 \AA, and we
include
the Galactic extinction
law of Cardelli, Clayton \& Mathis (1989) with R$_{V}=3.1$ and
E(B-V)=0.069 (Schlegel, Finkbeiner, \& Davis 1998) in the continuum fits.
The best fit parameters are $\alpha=1.082\pm0.021$ and 
$f_{1000}=8.18\pm0.06 \times 10^{-14}$ ergs s$^{-1}$ cm$^{-2}$ \AA$^{-1}$.   
For the 2004 \stis\
spectrum, we fit $f_{\lambda} = 3.25\pm0.01 \times 10^{-14} 
\lambda^{-1.086\pm0.007}$ ergs s$^{-1}$ cm$^{-2}$ \AA$^{-1}$.
We fit Gaussian profiles to the emission lines in all the spectra,
the most prominent of which are broad and narrow components of  
\ion{O}{6} in the \fuse\ band, and Ly$\alpha$ and \ion{C}{4}
in the \stis\ band.  The Ly$\alpha$ profile requires four
Gaussian components for a reasonable fit- broad, intermediate and 
two narrow components.   The \ion{C}{4} emission line requires
one broad, one intermediate, and one narrow component for a good fit, 
and we fit intermediate width and narrow 
components to the \ion{N}{5} emission.
These emission lines are plotted in 
Figures~\ref{fig-fusespec}  and \ref{fig-stisspec},
and the fit parameters are tabulated in
Table~\ref{table-emspec}.

\subsubsection{Interstellar and Intergalactic Absorption Features}
\label{sec-ism}

There are 4 velocity components of 21-cm emission along the 
\ngc\ sightline, including
one high velocity component associated
with the Magellanic Stream at $-333$~\kms (Wakker et al.\ 2001).
The strongest component of the interstellar absorption, 
with strong molecular hydrogen
absorption (Wakker et al.\ 2003), lies at zero velocity.  
We divide the ISM absorption from the \fuse\ spectrum using 
a two-temperature 
model of the zero velocity absorption constructed from the \fuse\
spectral simulation code FSIM.
Because the \stis\ spectra are much less crowded, 
we can use an empirical approach, modeling the ISM lines in the 
vicinity of intrinsic lines by fitting unblended lines of the same species
and scaling by the appropriate oscillator strength ratios from published atomic
data.
 
We identify an intergalactic absorption feature at $z=0.0134$ from its 
strong Ly$\alpha$ line in the \stis\ spectra.  
The qualitative shape of this feature is not as broad and shallow as the
feature at 1419 \AA\ caused by the repeller wire shadow discussed above,
and no such feature is expected at this position, or
flagged near this position in the data quality array.

Curiously, this intergalactic
Ly$\alpha$ lies between the two components of intrinsic 
absorption we discuss below.
We consider this system intergalactic for the following reasons:
(1) We find no  \ion{O}{6}, \ion{N}{5}, or \ion{C}{4} absorption
corresponding to this system; and 
(2) The absorption does not vary between the 2002 and 2004 observations
as the UV flux decreases by a factor of 2.5.
This could because its absorption is highly saturated, but we
rule out this possibility, even for a small covering fraction because no 
corresponding Ly$\beta$ absorption is seen in the \fuse\ spectrum.
Fixing the covering fraction to unity, we use {\it specfit} to derive
$N($\ion{H}{1}$)=6.7 \pm 0.3 \times 10^{12}$ cm$^{-2}$ and $b=52\pm2$ \kms\
for this intervening absorber.

\subsubsection{Intrinsic Absorption Features}
\label{sec-intabs}

Kriss et al.\ (2003) reported two primary components of UV absorption,
labeled Components 1 and 2, each with several velocity subcomponents.
We use the IRAF task {\it specfit} to 
fit the absorption profiles as Gaussians in optical
depth, specified by four parameters:  (1) the optical depth at line
center, (2) the line centroid, (3) the line width, and (4) the
covering fraction.  For doublets and members of a line
series, we fix the central optical depths to the ratios
required by the products of the wavelengths and the oscillator strengths. 
From these fits, we  derive column densities and
covering fractions for all subcomponents of the absorption complexes,
Components 1 and 2.  The errors on these parameters
are determined by {\it specfit} by
evaluating  and inverting the  curvature  matrix 
around the final value  of $\chi^{2}$.
For the fits to the absorption profiles, we fix the
continuum and emission line fits to those derived from
unabsorbed regions, as described above.

Analysis of high S/N spectra by 
Arav et al.\ (2002), Gabel et al.\ (2003, 2005), and 
Scott et al.\ (2004) shows that the profiles of intrinsic absorption lines in
AGN are best described using a velocity-dependent covering fraction. The \fuse\
and \stis\ data for \ngc, however, are too noisy for a reliable application
of this technique. In addition, because we lack the full Lyman series from the
\fuse\ data and because the \ion{C}{4} and \ion{N}{5} doublets are weak in
Component 1, as we discuss further below, we do not attempt such a detailed
analysis of these data. The next-best approach, as used by Kriss et al.\
(2003), decomposes the line profile with multiple blended Gaussians. Properly
fixed ratios of optical depth in line doublets enables us to constrain the
covering fraction in the individual components. Kriss et al.\ (2003) found that
the quality of the \fuse\ data were such that a single value of the covering
fraction adequately described each subcomponent. For consistency, and because
our data have even lower S/N, we adopt the same approach here.

\begin{figure}
\epsscale{1.0}
\plotone{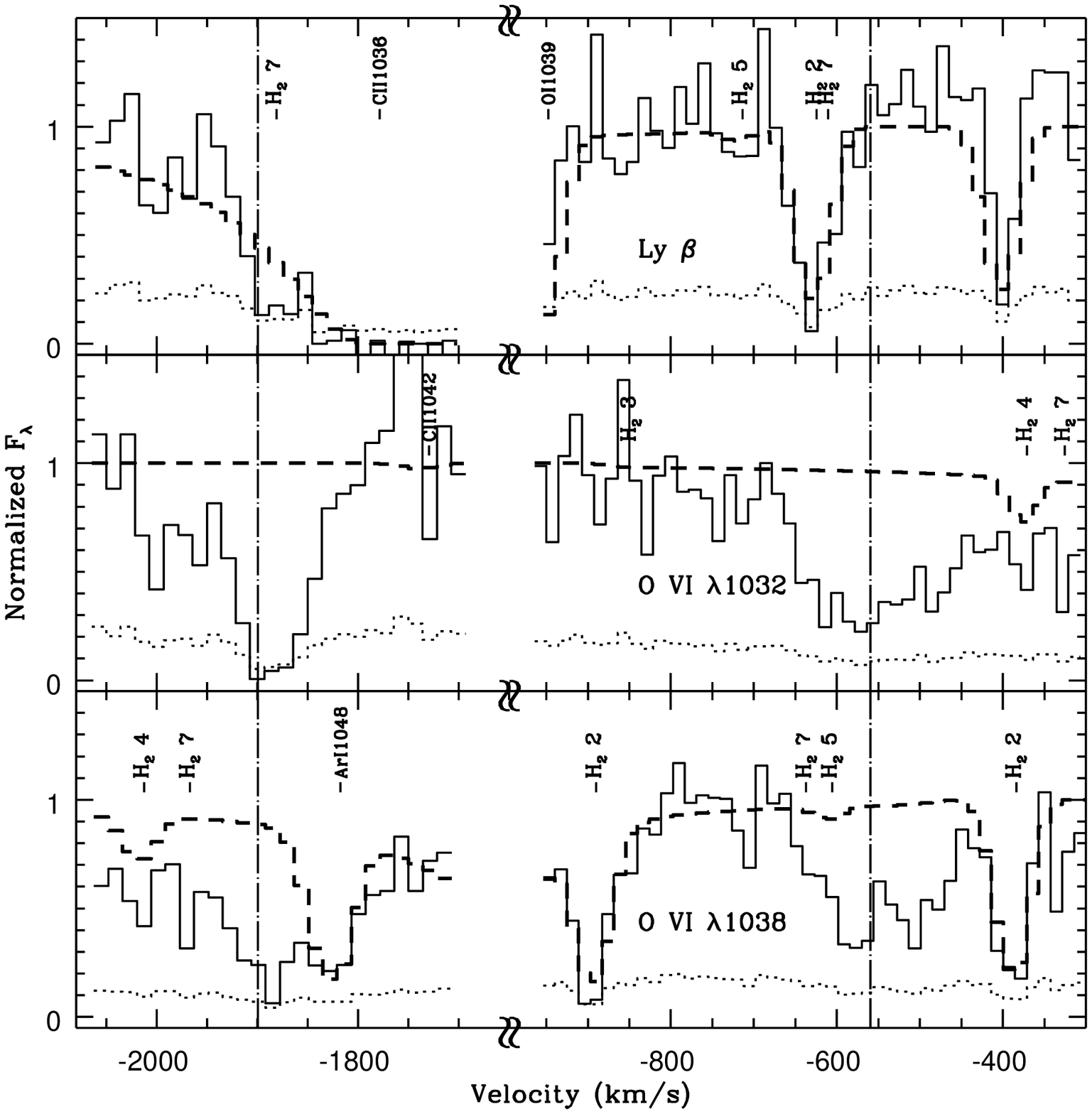}
\caption{Normalized 2002 \fuse\ spectrum shifted to
rest frame of \ngc,  with ISM fits (dashed lines) and 
Ly$\beta$ and \ion{O}{6} intrinsic absorption.  Vertical dot-dashed
lines represent the column density-weighted mean velocity
of all \ion{H}{1} subcomponents in Components 1 and 2,
$-562 \pm 11$ \kms\ and $-1901 \pm 3$ \kms, respectively.
\label{fig-fuseabs}}
\end{figure}

\begin{figure}
\epsscale{1.0}
\plotone{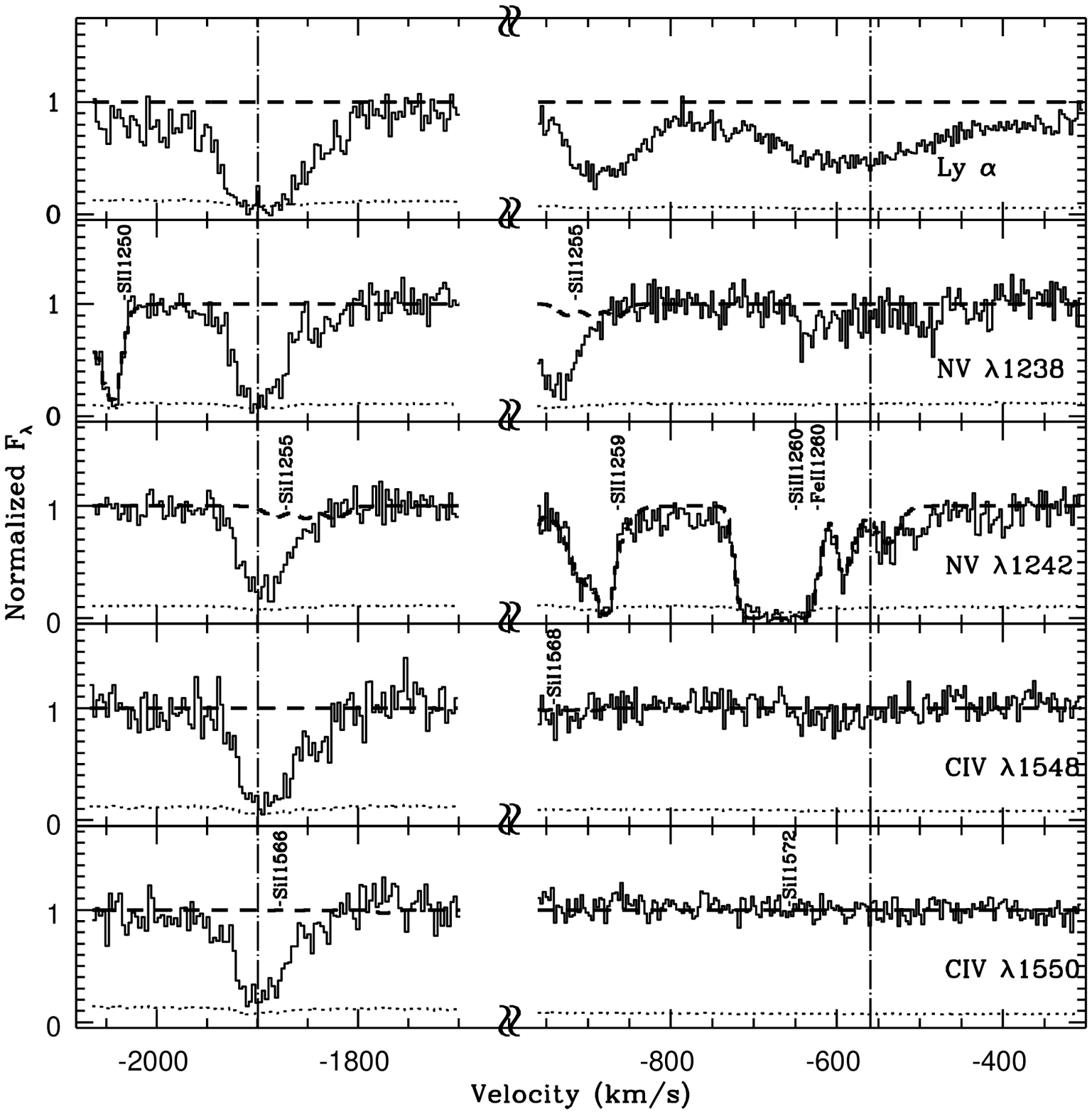}
\caption{Normalized 2002 \stis\ spectrum shifted to
rest frame of \ngc,  with ISM fits (dashed lines) and 
Ly$\alpha$, \ion{N}{5} and \ion{C}{4} intrinsic absorption.
Vertical dot-dashed
lines represent the column density-weighted mean velocity
of all \ion{H}{1} subcomponents in Components 1 and 2,
$-562 \pm 11$ \kms\ and $-1901 \pm 3$ \kms, respectively.
\label{fig-stisabs2002}}
\end{figure}

\begin{figure}
\epsscale{1.0}
\plotone{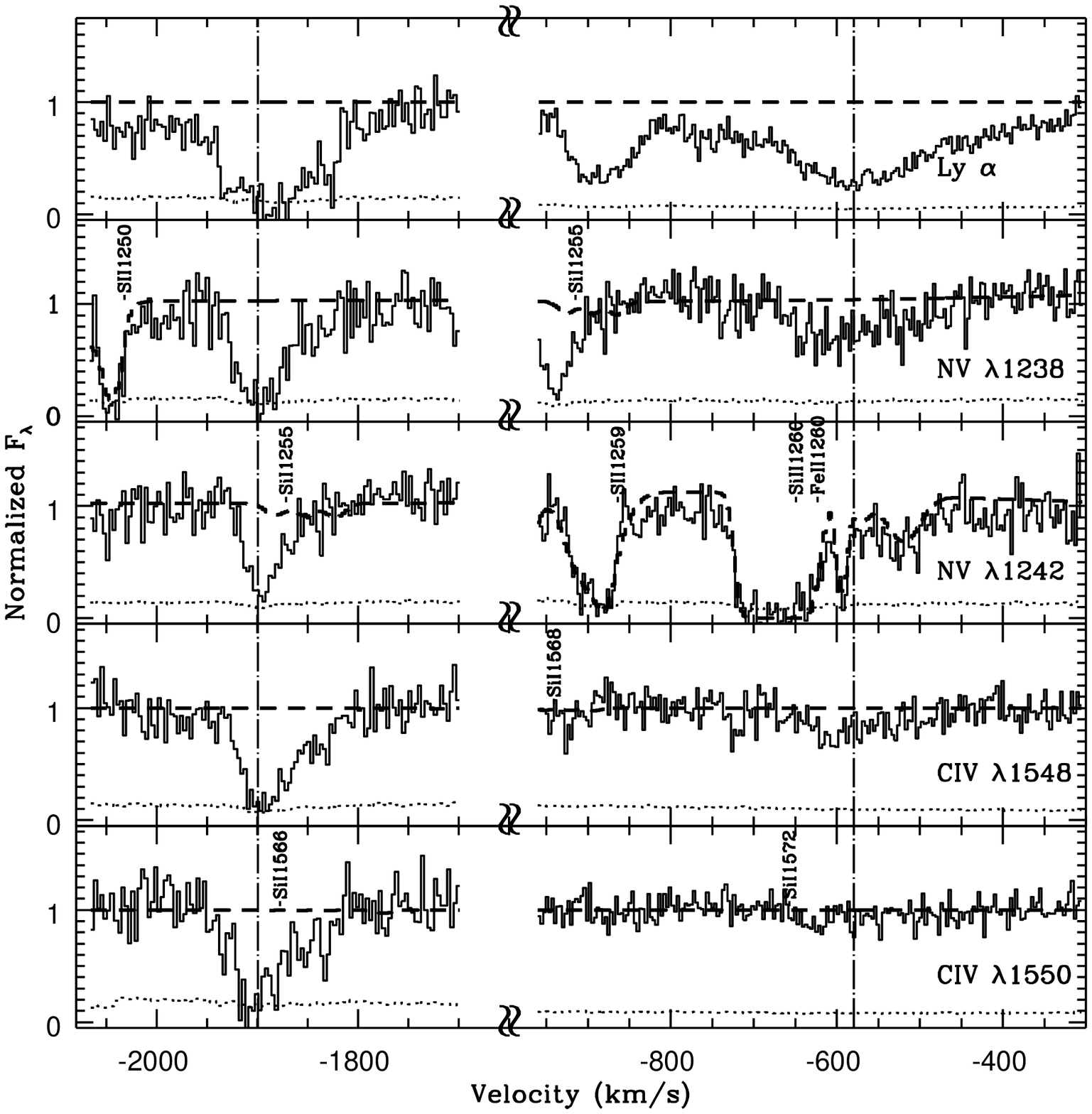}
\caption{Normalized 2004 \stis\ spectrum shifted to
rest frame of \ngc, with ISM fits (dashed lines) and 
Ly$\alpha$, \ion{N}{5} and \ion{C}{4} intrinsic absorption.
Vertical dot-dashed
lines represent the column density-weighted mean velocity
of all \ion{H}{1} subcomponents in Components 1 and 2,
$-583 \pm 9$ \kms\ and $-1902 \pm 3$ \kms, respectively.
\label{fig-stisabs2004}}
\end{figure}

In Component 2, we fit four subcomponents as
did Kriss et al.\ (2003).  However, in Component 1, we find blue and red
wings of absorption in Ly$\alpha$, leading to a fit consisting of
nine subcomponents.
We found that our fits did not require any change in the 
covering fraction of the absorption in either component 
in either the 2002 or the 2004 epoch observations.
We therefore fix the covering fractions to the 1999 values,
0.53 for Component 1 and 0.93 for Component 2.  We fix the  width
of all subcomponents at $61\pm4$ \kms\ and $48\pm5$ \kms\ (FWHM) for 
Components 1 and 2, respectively.
The resulting outflow velocities and column densities in all the
absorption subcomponents are listed in Table~\ref{table-abs}.
For the purposes of our analysis, we use the summed  column
densities of all the subcomponents in Components 1 and 2.  
The $N($\ion{H}{1}$)$-weighted outflow velocity of
Component 1  is $-562 \pm 11$ \kms\ and that of Component 2
is $-1901 \pm 3$ \kms.
These are plotted in Figures~\ref{fig-fuseabs}, 
\ref{fig-stisabs2002}, and \ref{fig-stisabs2004}
for all the species identified in intrinsic absorption in the \fuse\ and
\stis\ spectra.  

The \ion{H}{1} absorption is identified solely by the 
presence of the strong Ly$\alpha$ lines in the \stis\ spectrum,
seen in the top panels of Figures~\ref{fig-stisabs2002} and
\ref{fig-stisabs2004}.  In the 2002 \fuse\ spectrum 
no significant Ly$\beta$ absorption is detected.  For Component 1,
this is consistent with the low \ion{H}{1} covering fraction of the 
system.
The  Ly$\beta$  in Component 2 lies at the position of a strong
interstellar \ion{C}{2} line.  See Figure~\ref{fig-fuseabs}.

\begin{figure}
\epsscale{1.0}
\plotone{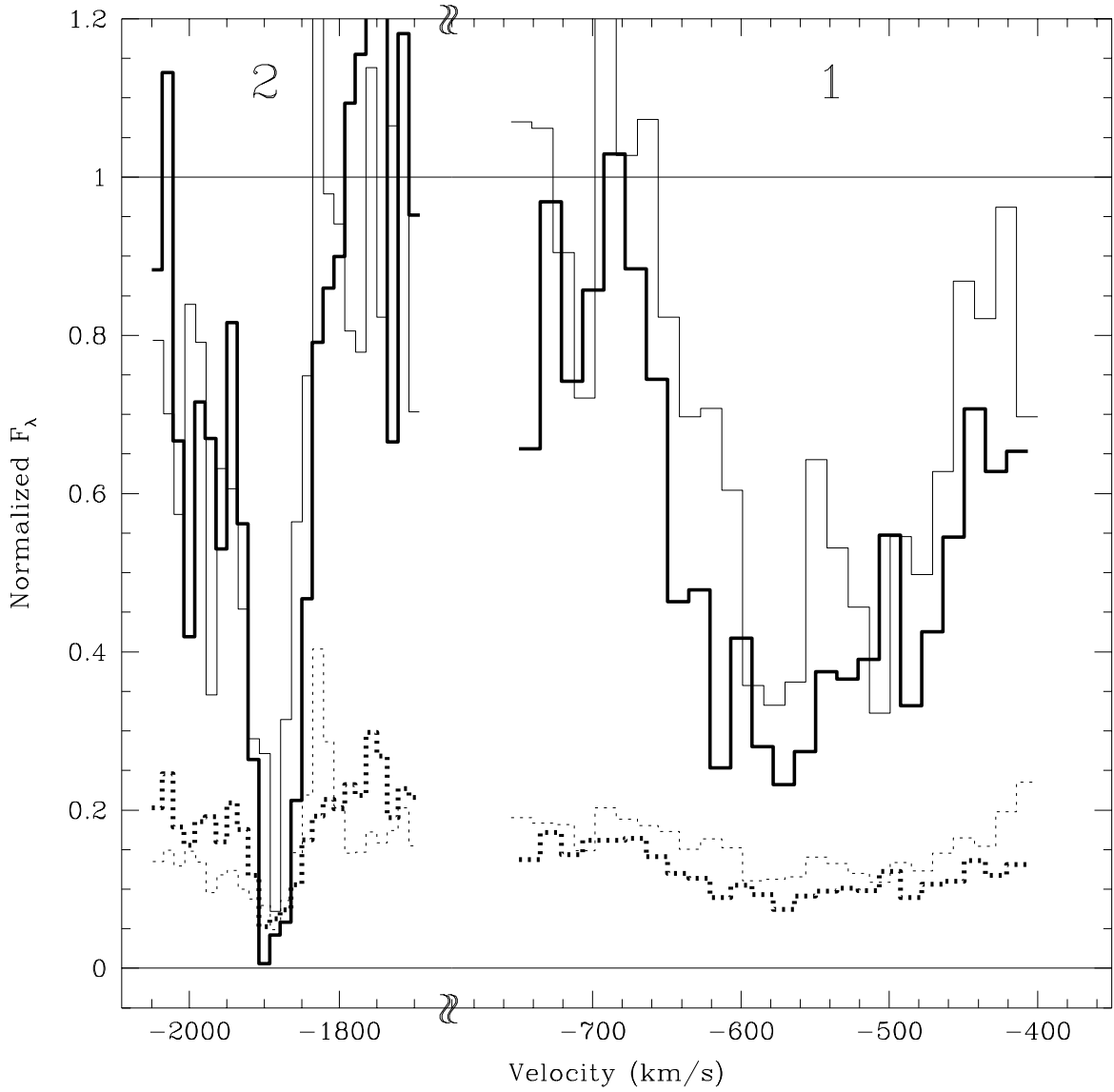}
\caption{\ion{O}{6} doublet in Components 1 and 2.  The
flux in the 1032 \AA\ line is shown by the solid bold lines,
errors are shown by the dotted bold lines.  The 1036 \AA\ lines
are shown by the regular solid and dotted lines.
\label{fig-o6sat}}
\end{figure}

The \ion{O}{6} doublet is strong in both components.  
As noted by Kriss et al.\ (2003), the strengths of the
doublet lines in Component 1 are equal, or nearly so
as best as can be determined by the signal-to-noise ratio
in the 2002 spectrum, which results from an exposure $\sim$5.5
times shorter than the 1999 spectrum.
Figure~\ref{fig-o6sat} shows both lines of the \ion{O}{6} doublet for
Components 1 and 2.    It is clear from this figure that in
Component 1, the flux ratio in the two line profiles 
is consistent with one within
the uncertainties.  The same is true, however, for 
Component 2 in this spectrum. 
We performed the initial fit the \ion{O}{6} absorption in Component 1 by
fixing the velocity offsets and column density ratios
of all the subcomponents with respect to the strongest subcomponent,
1d, to those measured for the strong \lya\ absorption in the \stis\ spectrum.
We found that allowing the velocity offsets of the subcomponents to
vary from the fiducial \lya\ values did reduce the $\chi^{2}$ of the
fit enough to justify the extra degrees of freedom, and the values
in Table~\ref{table-abs} reflect the improved fit.
Though we are able to measure the \ion{O}{6} column density in
Component 1 formally, by summing all the subcomponents, we
treat the measurement as a lower limit, since the higher S/N
1999 \fuse\ spectrum showed that this component is highly saturated.
This limit is shown in Figure~\ref{fig-sum}, which we discuss
further below in Section~\ref{sec-var}.
We note the possibility that the \ion{O}{6} column density is different
in these 2002 observations than in the 2000 epoch, because the 
UV continuum level has decreased and because 
$N($\ion{H}{1}$)$ increased from its 1999 value.  We
see no evidence for damping wings in the \ion{O}{6} profile, so 
an increase in $N($\ion{O}{6}$)$ from 1999 would imply that the
line is only further saturated and our
ability to measure the column density precisely has not improved.  
We do not assume saturation 
for Component 2, because in several velocity bins,
the flux ratio between the blue and red profiles 
is significantly greater than one, and again, because the 
1999 spectrum indicates that the absorption is only nearly,
not fully saturated.

We identify prominent \ion{N}{5} and \ion{C}{4} doublets 
associated with Component 2 in the \stis\ spectra.
These species, including Ly$\alpha$ absorption, are all
identified in the 1996 epoch data from the
{\it HST}/Faint Object Spectrograph and 
from the {\it International Ultraviolet Explorer}
presented  by Kriss et al.\ (2000).
However, we also identify weak but significant
\ion{N}{5} and \ion{C}{4} doublets associated with
Component 1.   These are shown in Figures~\ref{fig-stisabs2002} and
\ref{fig-stisabs2004}.  While the red lines of
each doublet are not obviously present in the spectrum, the blue members
of the doubles are visible in at least some subcomponents.  
We established the significance of
these doublets by comparing the $\chi^{2}$ of a fit with no
\ion{N}{5} and \ion{C}{4} to that of a fit
which did include these doublets. 
We fit these systems by fixing the covering fractions and
line widths
as discussed above.  We allowed the column density and
line center of subcomponent 1d to vary and 
fixed the velocity offsets and relative column densities 
of all the other other subcomponents
to those of the \ion{O}{6} subcomponents. 
For \ion{N}{5}, the improvement in the $\chi^{2}$ is
dramatic in both the 2002 and 2004 \stis\ spectra:
$\Delta\chi^{2}=152$ and $\Delta\chi^{2}=343$, respectively,  for
the extra 2 degrees of freedom. 
For \ion{C}{4} we find $\Delta \chi^{2}=22$ and
$\Delta \chi^{2}=201$.  From this we conclude that these
absorption features are indeed present in the data.
The results of all these absorption profile fits
are listed in Table~\ref{table-abs} and the summed column densities
are plotted in 
Figure~\ref{fig-sum}.

\subsubsection{Variability}
\label{sec-var}

Using previous FOS and \stis\ data along with the data presented here, 
we plot the variation of the 1000 \AA\ flux and the UV spectral
index in Figure~\ref{fig-var}.
From the 1996 observation to 2002, the 1000 \AA\ flux
increased by 26\% while the power law index became harder by
a similar factor.
Between 2000 and 2002, the 1000 \AA\ flux decreased by a factor
of $\sim$1.6, and by another factor of 2.5 between 2002 and 2004.
During this time, the spectral index got marginally softer, as
expected, but the change is not significant given the uncertainties.

The intrinsic Ly$\alpha$ and \ion{C}{4} $\lambda$1548 
absorption in the 2004  \stis\ spectrum is
shown in comparison with that in the 2002 spectrum in 
Figure~\ref{fig-comp}.
We plot the total \ion{H}{1}, \ion{O}{6},
\ion{C}{4}, and \ion{N}{5} column densities in
Components 1 and 2 in Figure~\ref{fig-sum}.
The top panel of the figure includes the upper limit on
$N($\ion{H}{1}$)$ from the nondetection of Ly$\beta$ in the 
1999 \fuse\ spectrum presented in Kriss et al.\ (2003), as well
as the measurements from Ly$\alpha$ in the 2002 and 2004 \stis\
spectra.  Also, in the second panel, we show the
$N($\ion{O}{6}$)$ measurements for Component 2 and the
lower limits  from the saturated absorption in Component 1
in the 1999 and 2002 \fuse\ spectra.

From Figures~\ref{fig-comp} and \ref{fig-sum}, we see that
the absorption in Component 2 does not vary in most species,
with the exception of an increase in $N($\ion{H}{1}$)$ between
1999 and the 2002/2004 epochs.
In Component 1, the \ion{H}{1} absorption grows stronger, consistent with
a change in the column density of a factor of 
$\sim4$ from 1999 to 2002 
and again by another factor of 2.3 from 2002 to 2004,
as the UV flux decreased by a total of a factor of
4.  The lower limit on $N($\ion{O}{6}$)$ is consistent with no change
since the 1999 observation.  The \ion{C}{4} column density
increased by a factor of 4 between the 2002 and 2004 observations, and
the \ion{N}{5} column density also increased by a smaller factor 
as the UV flux decreased by a factor of 2.5.  Thus, both
velocity components of the intrinsic absorption respond over time 
to changes in the ionizing flux.   In Section~\ref{sec-disc},
we discuss how we use this variability to place limits
on the radial distances of the absorbers from the AGN.

\section{Photoionization Models}
\label{sec-photo}
We computed photoionization models for the absorbing gas similar
to those of Krolik \& Kriss (1995, 2001) using the
XSTAR photoionization code  (Kallman et al.\ 2000).
In Figure~\ref{fig-sed}, we show the source spectral energy 
distribution used in the models.  For SED1, we used the
UV and X-ray spectral slopes and normalizations fit
to the 2002 epoch spectra of \ngc, described above in 
Sections~\ref{sec-xray} and \ref{sec-uvspec}. 
For comparison, we show SED2, the SED based on the 2000
epoch \fuse\ and \stis\ observations of \ngc,
presented used by Kriss et al.\ (2003).
As described by Kriss et al.\ (2003),  we assume
$f_{\nu} \propto \nu^{-1}$ at long wavelengths for both
SED1 and SED2.  This spectrum breaks at 2500 \AA\ to
$f_{\nu} \propto \nu^{-0.92}$ for SED1 and 
$f_{\nu} \propto \nu^{-0.75}$ for SED2 to match the
2000 and 2002 epoch \fuse\
spectra, respectively.  
The X-ray data are matched by 
$f_{\nu} \propto \nu^{-0.8}$ for SED1
$f_{\nu} \propto \nu^{-0.7}$ for SED2 between
0.5 keV and 100 keV.

We ran models in two two-dimensional grids of total absorber column density
and ionization parameter, appropriate for Components
1 and 2:  ($\log N$, $\log U$) = (19.5-20.75, $-3.0$-4.0) and
(18.3-19.0, $-5.3$-$-0.3$), respectively, assuming solar abundances
(Grevesse, Noels, \& Sauval 1996)
in both cases.
We then compared the
column densities of the observed species:
\ion{H}{1}, \ion{O}{6},
\ion{N}{5}, \ion{C}{4} in the UV data, and  the H-like and He-like
Ne, Mg, and Si observed in the \chand\ data, as well as limits 
on \ion{O}{7} and \ion{O}{8}
to the values predicted by the models
in order to determine the properties of the two velocity components of 
the intrinsic absorption. 

For Component 1, we find that $U=1.0$  is
consistent with the measurements of \ion{N}{5} and \ion{C}{4}, with
the lower limit on the saturated \ion{O}{6}
absorption, and with the 
upper limits on the \ion{O}{7} and \ion{O}{8} absorption
from their nondetection in the \chand\ data, for our 
assumption of $b=100$ \kms, but also for Doppler
parameters as large as $\sim1000$ \kms.
For the total column density, we find $\log N= 20.0$ is a 
reasonable match to the absorbing columns.
For Component 2, we find ($\log N$, $U$) = (18.6, $0.08$) 
gives the best match with the absorption data.

All of the absorption from H-like and He-like ions
in the \chand\ data is seen at
velocities consistent with the outflow velocity of Component 1.
However, the column densities we measure imply larger ionization parameters
from the models, $8 \lesssim U \lesssim  12$,
implying $\log N > 20$, indicating the
presence of multiple ionization phases, as found in the
warm absorber associated with NGC~3783 (Netzer et al.\ 2003; Krongold
et al.\ 2003) from spectral fitting to photoionization models. 
The relative 
dearth of spectral features in our \chand\ spectrum of NGC~7469
precludes such detailed modeling in this case.

\section{Discussion}
\label{sec-disc}

These new simultaneous UV and X-ray observations of
\ngc\ allow us to compare the column densities of various
species observed in the absorbers
with photoionization models in a fully self-consistent manner
in order to infer the physical conditions in the outflow.
From our multi-epoch UV observations, we use the variability
of the ionizing continuum and the absorption to
draw conclusions about the distances of the absorbing components
from the central engine of the AGN.

Kriss et al.\ (2003) reported $U=6.0$ for
Component 1.
Although this value is consistent with
our measurement of \ion{O}{6} in the  2002 \fuse\ data, and the limits on
\ion{O}{7} and \ion{O}{8} from the \chand\ spectrum,
it is incompatible with the
detection of any \ion{N}{5} and \ion{C}{4} absorption in the
2002 \stis\ spectrum.  With the added leverage 
afforded by the column density
measurements of these species, we now estimate $U=1.0$ 
for this velocity component. 
This decrease in the ionization parameter is consistent with
the decrease in the ionizing flux between the 1999 and
2002 observations, although 
$U=1$ is somewhat smaller than expected for a constant
total column density and a scaling of the number of
ionizing photons, which decreases by a factor of
$\sim$2 for the SEDs assumed for the 1999 and 2002 epochs
shown in Figure~\ref{fig-sed}.
The changes in 
the \ion{H}{1}, \ion{C}{4}, and \ion{N}{5}
column densities between the \stis\ 
observations in 2002 and 2004 are also
consistent with photoionization changes in the
absorber.   We have no knowledge of the X-ray continuum
at the 2004 epoch, but if we make the simple assumption
that the X-ray flux
decreases in direct proportion to the UV flux, the
column densities of these species are in
rough agreement with $\log N=20$ and $U=0.4$.  We
lack the constraining power of a measurement
of $N($\ion{O}{6}$)$ at this epoch.

As Steenbrugge et al.\ (2003) found for the warm absorber in
NGC~5548 and Ogle et al.\ (2004) found for the warm absorber in
NGC~4051, we find
a trend of increasing column density with increasing ionization parameter
for the species we detect in Component 1 of the \ngc\ absorber.
In Figure~\ref{fig-peak} we show the ionization parameter
at which the abundance of each species peaks
versus the equivalent hydrogen column
density for that species in Component 1 with a least-squares estimation
of the power law fit.

The total hydrogen column densities derived from the \ion{Fe}{20} and 
\ion{Fe}{21}
lines would imply that the absorber is Compton thick.  These columns are
derived, as all the points in Figure~\ref{fig-peak} are, 
from the measured line
equivalent widths using the curve of growth and a Doppler parameter of 
100 \kms.  We use the
fractional abundance of the ion from the photoionization models and the
assumed solar abundance of each atomic species.  There are two ways the
points could move down on Figure 19:  (1) a larger Doppler parameter, eg.
$b=500$ \kms would yield 
$N_{\rm H}($\ion{Fe}{20}$)= 2 \times 10^{23}$~cm$^{-2}$ and 
$N_{\rm H}($\ion{Fe}{21}$)= 3 \times 10^{24}$~cm$^{-2}$,
however, we note that this would affect all the points in the
figure;
or (2) an enhanced iron abundance in the absorber, as seen in some
AGN broad line regions (Hamann \& Ferland 1999) and accretion disks (Lee et
al.\ 1999, Ballantyne \& Fabian 2001, Bianchi \& Matt 2002, Schurch et
al.\ 2003).  
In any case, removing 
the \ion{Fe}{20} and \ion{Fe}{21} points from the fit does 
not change the slope
or the standard deviation of the regression significantly.

To treat the lower limit on $N_{\rm H}($\ion{O}{6}$)$
and the upper limits on $N_{\rm H}($\ion{O}{7}$)$ and 
$N_{\rm H}($\ion{O}{8}$)$
we used the IRAF {\it emmethod} algorithm appropriate for censored data
to perform the linear regression.
In order to determine if the regression solution is driven primarily by
the low-U points in Figure~\ref{fig-peak},
we performed the linear fit without the
\ion{C}{4}, \ion{N}{5}, and \ion{O}{6}
points included. 
The standard deviation of the fit and of
the individual fit parameters increases, but within this increased
uncertainty, the changes in the fit parameters are not significant.
This fit is shown by the dotted line in Figure~\ref{fig-peak}.
This relationship between $U$ and $N_{\rm H}$ illustrated  in
Figure~\ref{fig-peak} indicates that, while 
$U=1.0$ for Component 1 is broadly consistent with the UV data,
the absorber is composed of gas with a 
range of ionization parameters and the absorbing
column scales with ionization. 

\begin{figure}
\epsscale{1.0}
\plotone{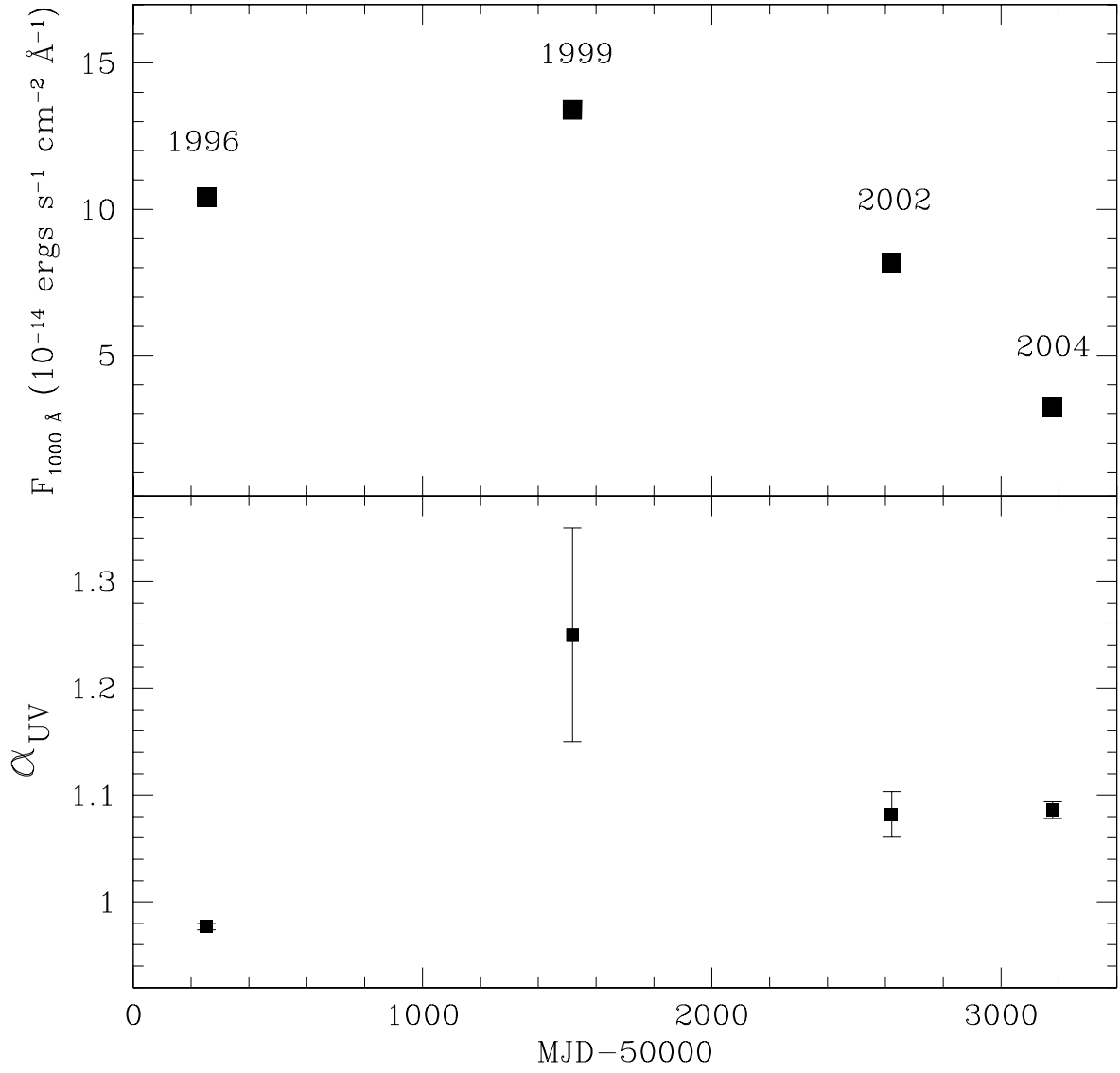}
\caption{Variation in the continuum flux and the UV spectral index of NGC~7469.
\label{fig-var}}
\end{figure}

\begin{figure}
\epsscale{1.0}
\plotone{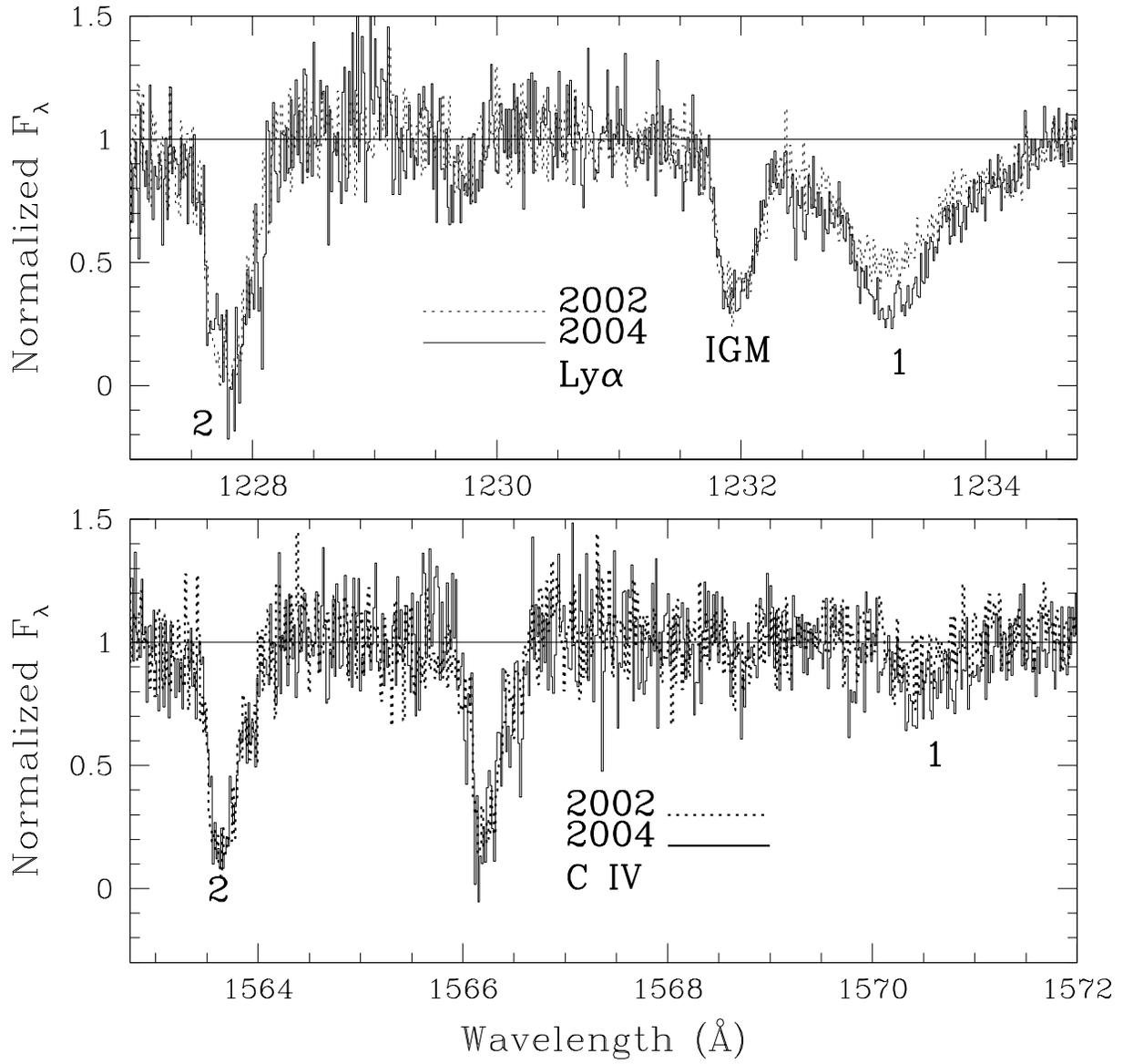}
\caption{Comparison of absorption in Ly$\alpha$ and \ion{C}{4} between 2002 and 2004.
\label{fig-comp}}
\end{figure}

\begin{figure}
\epsscale{1.0}
\plotone{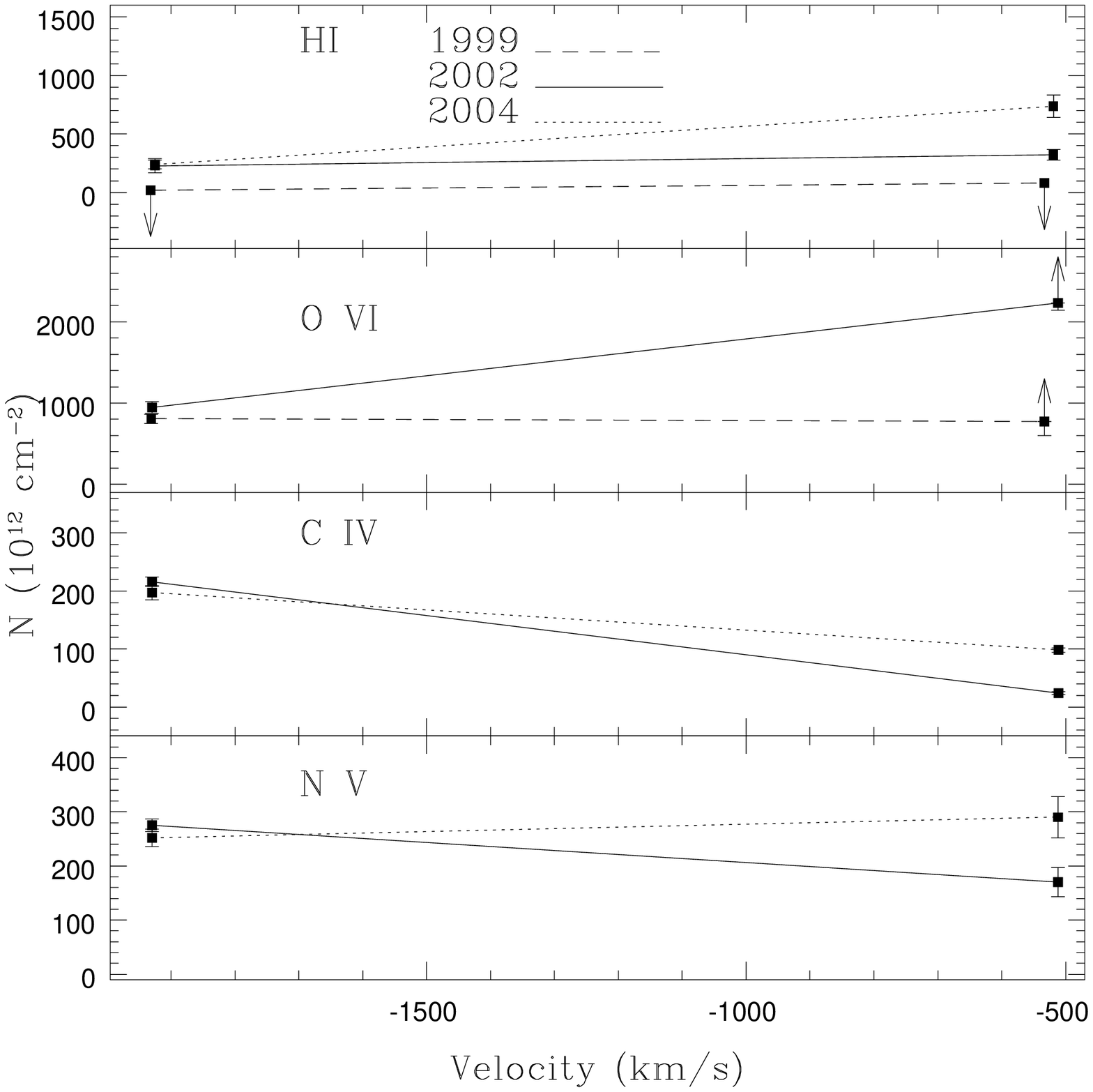}
\caption{Summary of \ion{H}{1}, \ion{O}{6}, \ion{C}{4}, and \ion{N}{5} 
absorption intrinsic to \ngc\ in 1999, 2002, and 2004 epochs.
\label{fig-sum}}
\end{figure}

\begin{figure}
\epsscale{1.0}
\plotone{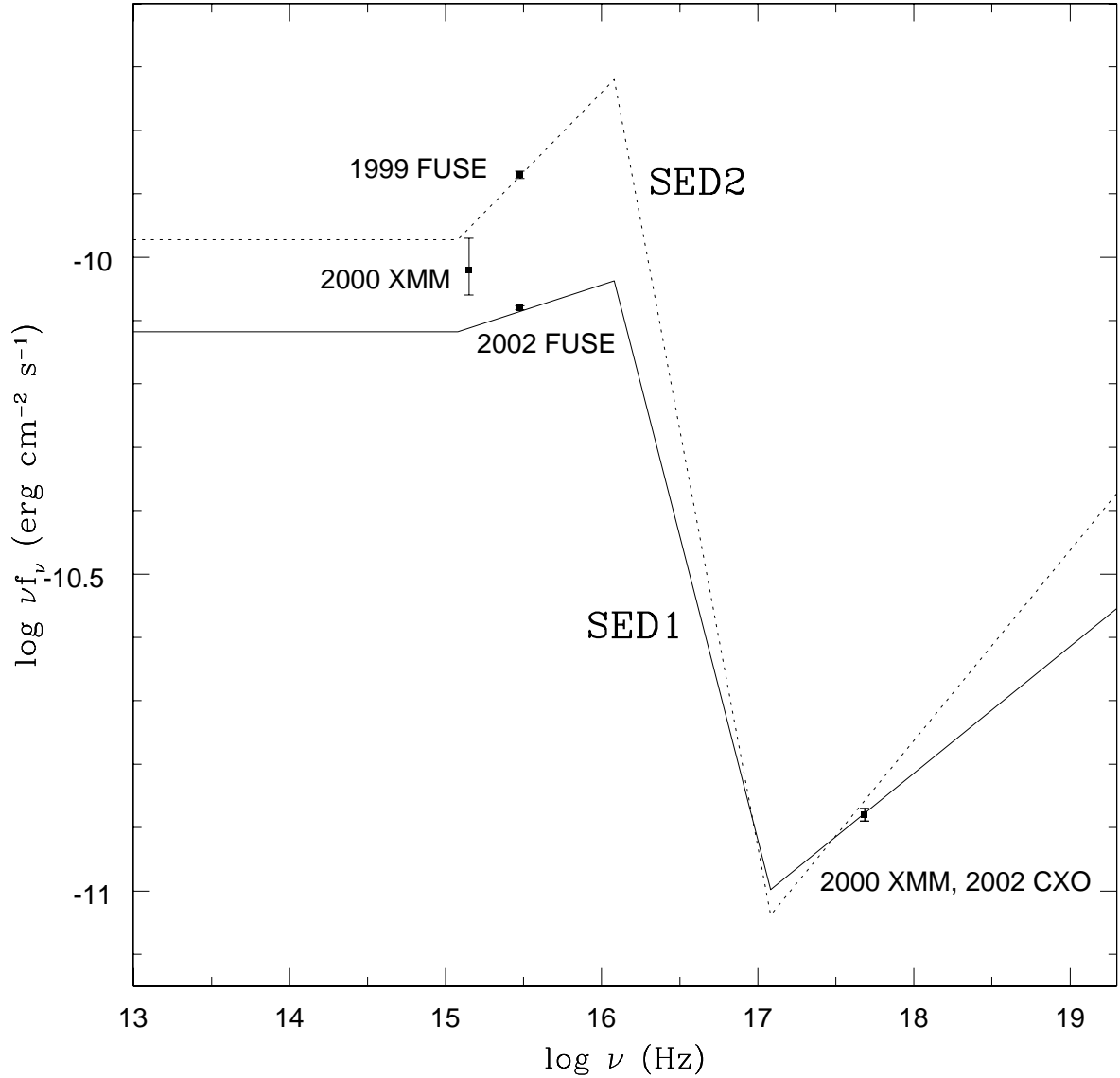}
\caption{Spectral energy distributions used in photoionization
models of the intrinsic absorbers in \ngc. SED1 is normalized
such that it matches the 1000 \AA\ flux in the 2002
epoch \fuse\ spectrum and the 2 keV flux in the \chand\ spectrum.
SED2 is normalized to the 2000 epoch \fuse\ and {\it XMM-Newton}
data plotted, as described by
Kriss et al.\ (2003).}
\label{fig-sed}
\end{figure}

\begin{figure}
\epsscale{1.0}
\plotone{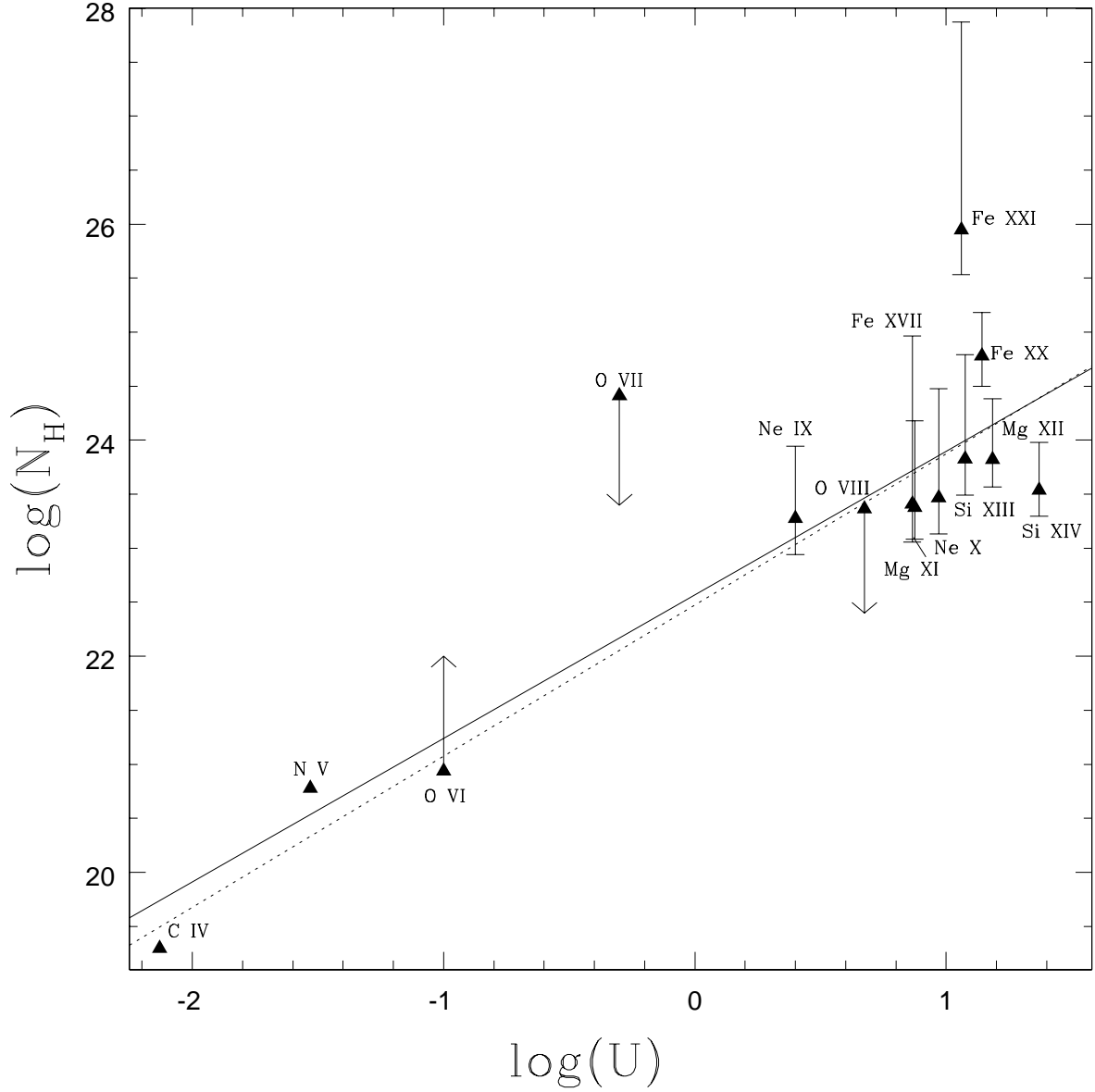}
\caption{Equivalent hydrogen column density versus 
ionization parameter at peak abundance of each ion with power law
fit. The dotted line shows a fit with the \ion{C}{4}, \ion{N}{5},
and \ion{O}{6} points omitted.}
\label{fig-peak}
\end{figure}

The column density and ionization parameter we infer for
Component 2 are similar to the values inferred by
Kriss et al.\ (2003) for the 1996 FOS and {\it IUE} observations
(Kriss et al.\ 2000a).  This is consistent with a change
in the ionization parameter due to a change in the ionizing flux,
as the number of ionizing photons from SED1 is comparable
to the number inferred from the Kriss et al.\ (2000a) SED.
(See Figures~\ref{fig-var} and \ref{fig-sed}.)

We find X-ray absorption in helium-like and hydrogen-like
neon, magnesium, and silicon all at velocities consistent
UV Component 1 to within $2\sigma$,
although not with the same total column density and ionization
parameter as the UV-absorbing species.
This supports the 
conclusions of Kriss et al.\ (2003) and Blustin et al.\ (2003), 
that this component is associated with the
X-ray warm absorber and is located closer to the central
engine than Component 2.  This is further corroborated by the
small covering fraction and outflow velocity of Component 1, 
$\sim0.5$ and $-560$ \kms, respectively, compared
with the unity covering fraction and $\sim -1900$ \kms\ 
velocity of Component 2 with respect to the systemic
redshift of \ngc.  The small covering fraction of Component 1
may indicate full covering of the continuum source and  
partial covering of the BELR.
This geometry is consistent with the larger size of the BELR
with respect to the central, continuum emission source, and
has also been inferred in the outflow in Mrk~279
by applying simultaneous fits to several Lyman series lines and
to the \ion{C}{4}, \ion{N}{5}, and \ion{O}{6} doublets
(Scott et al.\ 2004; Gabel et al.\ 2005).
The increase in the
\ion{H}{1}, \ion{N}{5}, and \ion{C}{4} column densities 
in Component 1 with the decrease in the flux in the ionizing continuum
also supports this interpretation.  

The only X-ray absorption 
feature we find in common with the results of Blustin et al.\ (2003)
is the \ion{Ne}{9} He$\beta$ line at a rest wavelength of 
11.5 \AA\ identified in Component 1.   We report its
equivalent width as $14 \pm 5$ m\AA, while Blustin et al.\ (2003)
found $50 \pm 20$ m\AA. The marginal decrease in the
\ion{Ne}{9} absorption is consistent with a photoionization effect
and the adopted SEDs shown
in Figure~\ref{fig-sed} in the sense that the flux at the
ionization potential of \ion{Ne}{8} is 23\% larger in the 
2000 epoch SED (SED1) relative to the 2002 SED (SED2).
The total column density in Component 1 implied by UV absorption 
features from the photoionization modeling discussed in 
Section~\ref{sec-photo}, $\log N = 20$, is consistent with
that found for the X-ray absorber by Blustin et al.\ (2003) 
from their {\it XMM} data.  However, as noted above, and as
illustrated by Figure~\ref{fig-peak}, the
hydrogen-like and helium-like absorption we find in the
\chand\ data imply a higher total column density indicating
a range of ionization phases. 

We use the fact that we see variability in the absorption in Component 1,
particularly in  its \ion{H}{1} absorption, and the time between
the 2002 and 2004 observations to place a 540-day upper limit on
the recombination time of the gas, where 
$t_{rec} \equiv (n_e \alpha_{rec,i} \frac{n_{i+1}}{n_i})^{-1}$ 
(Krolik \& Kriss 1995). For $T = 3 \times 10^{4}$~K, the
hydrogen recombination rate is
$\sim2.3 \times 10^{-13}$~cm$^{3}$~s$^{-1}$.
This gives a fairly weak lower limit
on the density, $n_e \gtrsim 45$ cm$^{-3}$, assuming a neutral fraction of 
$5 \times 10^{-4}$. 
Using the ionization parameter $U=1.0$,
this gives an upper limit on the radial distance of Component 
1 of $\sim$100 pc.
Following the prescription of Blustin et al.\ (2005)
we estimate the maximum radial distance under the assumption that
the absorption arises in a region of width less than or equal to the
distance from the ionizing source: 
${\rm r} \leq L_{ion} C_{v}({\rm r})/(\xi N_H)$,
where $C_{v}({\rm r})$ is the volume filling factor.  For the parameters
of the X-ray absorption derived from the 2000 epoch \chand\ observation
of NGC~7469 (Blustin et al.\ 2003), Blustin et al.\ (2005) derive 
$C_{v}({\rm r})=0.086$\% and $r \leq 1.6$ pc for the X-ray absorber, which
we argue is the same as Component 1 of the UV absorption.
Using the parameters derived from the UV absorption, and
assuming the same volume filling factor as Blustin et al.\ (2005),
we find a somewhat firmer upper limit on the distance of Component 1
from the central engine than we could place from the
variability, ${\rm r} < 15$ pc.   If the UV absorption arises in
higher density material embedded in the X-ray warm absorber
(Krolik \& Kriss 1995, 2001), its
volume filling factor may be even smaller than that estimated above,
and the radial distance correspondingly smaller.

Blustin et al.\ (2005) found from their comprehensive analysis
of the warm absorber data in the literature that
most absorbers in Seyfert galaxies are located at
distances from the central engine consistent with the
torus, indicating that the outflow may originate there in
a scenario similar to that proposed by Krolik \& Kriss (1995, 2001).
In the case of \ngc, they found that the warm absorber
analyzed by Blustin et al.\ (2003) fits into this picture.
The minimum and maximum distances they estimate for the absorber
indicate that it lies outside the radius of the BELR,
$0.0042$ pc (Wandel, Peterson, \& Malkan 1999) and within
the radius of the torus, $\sim L_{44}^{0.5}$ pc (Krolik \& Kriss 2001)
where $L$ is the ionizing luminosity between 1 and 1000 Ryd.
Allowing for the lower ionizing luminosity relevant to the 2002
observations, a slightly larger column density weighted ionization parameter
for the X-ray absorption,
and a mean outflow velocity somewhat lower than
measured previously by Blustin et al.\ (2003), $U \sim 10$ versus
$U \sim 5$, and $-580$ \kms\
versus $-800$ \kms, we find a similar overall result.
For both the UV and X-ray absorption in Component 1,
we find ${\rm r}_{\rm min} >> {\rm r}_{\rm BELR}$, given
${\rm r}_{\rm min}= 2 G M_{BH}/v^{2}_{\rm out}$.
We note, however, that because the formal uncertainties on
the black hole mass from Wandel et al.\ (1999),
$M_{BH} = 0.76^{+0.75}_{-0.76} \times 10^{7}$ $M_{\sun}$,
allow the black hole mass to be as low as $1.5 \times 10^{5}$ $M_{\sun}$,
or less, making
${\rm r}_{\rm min}$ consistent with
the absorber arising from an accretion disk wind.
Blustin et al.\ (2005) propose the alternative explanation
that the wind has not yet achieved the escape speed
and may therefore be closer to the central
engine than ${\rm r}_{\rm min}$ as estimated above. 
This way, the partial covering of the continuum and/or 
broad line source demanded in particular 
by the saturated \ion{O}{6} profile of Component 1 may
be understood.   

The larger outflow velocity of
Component 2 compared to Component 1 
leads to a smaller lower limit on the distance of Component 2 
from the central engine, $\sim0.02$ pc.
However, given the unity covering fraction of this component, 
it is unlikely that
we can place it within the BELR,
even given the uncertainty in $M_{BH}$ noted above.  
We use the change in $N($\ion{H}{1}$)$ in Component 2 between the
1999 and 2002 epochs to estimate
its maximum distance from the ionizing source from setting
the recombination time to be less than the time between
these observations.
For a neutral fraction of
$5 \times 10^{-4}$, $t_{\rm rec} \leq 1100$ days implies
$n \gtrsim 20$ cm$^{-3}$.
The best-fit ionization parameter for this component, 
$U=0.08$, implies ${\rm r} <  600$ pc.

We conclude that there is still ambiguity in the
radial distances of the absorbers from the central engine
of \ngc,
but that the general picture described in previous work 
(Blustin et al.\ 2003; Kriss et al.\ 2003)
holds: Component 1 is a high
density, high ionization absorber located near or within
the BELR and Component 2 is an absorber of relatively lower
density and ionization located in a region that is 
consistent with an origin in the AGN torus.
 
\acknowledgements 
The authors thank K.\ Sembach for providing the 2004 epoch
\stis\ observations, A.\ Blustin for providing the 2000 epoch
{\it XMM}/RGS spectrum for comparison with our {\it Chandra} data, 
and T.\ Alexander for contributions to
the \fuse\ pipeline reductions.  We thank the anonymous
referee for useful comments.
J.\ E.\ S. acknowledges the support of a 
National Research Council Associateship held at NASA/Goddard
Space Flight Center. 
J.\ C.\ L. thanks and acknowledges support from the Chandra
fellowship grant PF2--30023, issued by the Chandra X-ray
Observatory Center, which is operated by SAO for and on behalf of NASA
under contract NAS8--39073.

\begin{deluxetable}{llccll}
\tablecolumns{5}
\tablewidth{27pc}
\tablecaption{Observations of NGC~7469 \label{table-obs}}
\tablehead{
\colhead{Instrument} &\colhead{ID} &\colhead{Start Date} &\colhead{UT} &\colhead{Exp. (s)} }
\startdata
{\it Chandra} &700395     &2002-12-12   &13:37:08  &79840 \\
{\it Chandra} &700586     &2002-12-13   &12:10:13  &69760 \\
{\it FUSE}    &C0900101   &2002-12-13   &07:07:39  &3596  \\
{\it FUSE}    &C0900102   &2002-12-13   &06:25:22  &3352  \\
STIS          &O6BN01010  &2002-12-13   &06:28:04  &2233  \\
STIS          &O6BN01020  &2002-12-13   &07:51:30  &2695  \\
STIS          &O6BN01030  &2002-12-13   &09:27:36  &2695  \\
STIS          &O6BN01040  &2002-12-13   &11:03:42  &2695  \\
STIS          &O6BN01050  &2002-12-13   &12:39:47  &2695  \\
STIS          &O8N501010  &2004-06-21   &17:55:10  &1940  \\
STIS          &O8N501020  &2004-06-21   &19:14:52  &2870  \\
STIS          &O8N501030  &2004-06-21   &20:50:51  &2290  \\
STIS          &O8N501040  &2004-06-21   &22:26:50  &2870  \\
STIS          &O8N501050  &2004-06-21   &00:02:50  &2870  \\
STIS          &O8N502010  &2004-06-22   &17:55:3   &1940  \\
STIS          &O8N502020  &2004-06-22   &19:14:49  &2290  \\
STIS          &O8N502030  &2004-06-22   &20:50:49  &2870  \\
STIS          &O8N502040  &2004-06-22   &22:26:49  &2870  \\
\enddata
\end{deluxetable}

\begin{deluxetable}{lcccccccc}
\tablecolumns{7}
\tablewidth{0pc}
\tablecaption{Spectral Features in {\it Chandra} Spectrum of NGC~7469
\label{table-cxo}}
\tablehead{
\colhead{$\lambda_{\rm vac}$} &\colhead{Feature}
&\colhead{Equiv.\ width} &\colhead{Flux\tablenotemark{1}}
&\colhead{Signif.}\tablenotemark{3}
&\colhead{$\Delta v$ \tablenotemark{3}} &\colhead{FWHM} \\
\colhead{} &\colhead{(\AA)} 
&\colhead{(m\AA)} &\colhead{} &\colhead{$\sigma$}
&\colhead{(\kms)} &\colhead{(\kms)} }
\startdata
\multicolumn{6}{c}{Absorption}\\
\hline
6.18  &Si~{\sc xiv}  Ly$\alpha$ &$5\pm2$  & &2.3 &$-540\pm260$ &$620\pm690$\\
6.65  &Si~{\sc xiii} He$\alpha$ &$11\pm6$ & &5.0 &$-600\pm330$ &$1350\pm900$ \\
8.42  &Mg~{\sc xii}  Ly$\alpha$ &$10\pm3$ & &2.6 &$-540\pm100$ &$500\pm160$\\
9.17  &Mg~{\sc xi}   He$\alpha$ &$11\pm5$ & &3.3 &$-620\pm130$ &$680\pm410$\\
11.55 &Ne~{\sc ix}   He$\beta$  &$14\pm5$ & &4.4 &$-910\pm180$ &$1020\pm520$ \\
12.13 &Ne~{\sc x}    Ly$\alpha$ &$17\pm7$ & &1.3\tablenotemark{4} &$-530\pm80$  &$430\pm270$ \\
12.28 &Fe~{\sc xxi}             &$21\pm11$& &2.2\tablenotemark{4} &$-540\pm240$ &$890\pm570$\\
12.82 &Fe~{\sc xx}              &$27\pm6$ & &5.0 &$-290\pm130$ &$1070\pm340$\\
13.45 &Ne~{\sc ix} He$\alpha$   &$24\pm9$ & &3.6 &$-670\pm180$ &$1050\pm490$ \\
13.45 &Ne~{\sc ix} He$\alpha$   &$25\pm6$ & &3.2 &$860\pm100$  &$760\pm220$ \\
15.01 &Fe~{\sc xvii}            &$19\pm8$ & &2.7 &$-690\pm150$ &$650\pm320$\\
\hline
\multicolumn{6}{c}{Emission}\\
\hline
      &Fe K$\alpha$ & &$3.9\pm0.7$ &17 &$6.39\pm0.01$\tablenotemark{5} 
        &$6310\pm1580$\\
2.7   &\nodata   & &$1.0\pm0.4$ &6.5 &0            &$3410\pm1600$\\
8.2   &\nodata   & &$0.8\pm0.4$ &6.2 &0            &$1600\pm940$\\
11.0  &\nodata   & &$1.1\pm0.5$ &4.5 &0            &$590\pm300$\\ 
12.13 &Ne~{\sc x} Ly$\alpha$      &    &$0.7\pm0.5$ 
        &1.3\tablenotemark{4}  &$430\pm170$  &$480\pm400$\\
13.2  &\nodata   & &$1.0\pm0.4$ &4.5 &0    &$790\pm480$\\ 
13.69 &Ne~{\sc ix}(f) & &$1.6\pm0.7$ &4.8 &$-200\pm290$ &$1250\pm690$\\
14.45 &O~{\sc viii} Ly-7?\tablenotemark{6} & &$1.0\pm0.4$ 
        &4.5 &$-180\pm170$ &$780\pm410$\\
14.52  &O~{\sc viii} Ly-6?\tablenotemark{6} & &$1.4\pm0.6$
        &5.9 &$810\pm280$  &$1270\pm660$\\
14.63  &O~{\sc viii} Ly-5?\tablenotemark{6} & &$1.5\pm0.8$
        &6.1 &$1470\pm470$ &$1790\pm1190$\\
15.4   &\nodata        & &$1.3\pm0.6$  &4.4 &0  &$820\pm450$\\  
15.5   &\nodata        & &$1.6\pm0.6$  &5.5 &0  &$790\pm340$\\
16.77  &O~{\sc vii} RRC& &$2.0\pm0.9$  &5.8 &$-1300\pm230$ &$1020\pm530$\\
17.1  &\nodata        & &$2.3\pm1.0$   &6.2 &0             &$1080\pm550$\\
17.3  &\nodata        & &$1.5\pm0.8$   &3.8 &0             &$720\pm460$\\
17.4  &\nodata        & &$1.4\pm0.7$   &3.1 &0             &$490\pm260$\\
18.3  &\nodata        & &$1.9\pm1.0$   &3.4 &0             &$700\pm410$\\
18.97 &O~{\sc viii} Ly$\alpha$ & &$3.0\pm1.0$  &5.1 &$50\pm90$    &$370\pm340$\\
20.2  &\nodata        & &2.4$\pm1.5$   &3.5 &0            &$620\pm440$\\
22.09 &O~{\sc vii}(f) & &$6.1\pm2.5$   &3.9 &$-200\pm90$  &$460\pm220$ \\
\enddata
\tablenotetext{1}{In units of $10^{-5}$ photons cm$^{-2}$ s$^{-1}$}
\tablenotetext{2}{Significance of feature from smoothed equivalent width
threshold. See \S~\ref{sec-xray}.}
\tablenotetext{3}{Velocity with respect to
systemic redshift of \ngc, $z=0.01639$ }
\tablenotetext{4}{\ion{Ne}{10} absorption and emission are detected in the MEG
spectrum at 4.1$\sigma$ and 3.6$\sigma$, respectively, and \ion{Fe}{21}
absorption is detected in the MEG spectrum at 3.4$\sigma$.}
\tablenotetext{5}{Rest-frame energy of profile center in keV}
\tablenotetext{6}{ID highly uncertain, no Ly$\beta$, Ly$\gamma$, or Ly$\delta$
detected. See Figure~\ref{fig-o8} and discussion in \S~\ref{sec-xray}.}
\end{deluxetable}

\begin{deluxetable}{llccc}
\tablecolumns{5}
\tablewidth{30pc}
\tablecaption{Emission Line Fits to {\it FUSE} and STIS Spectra of NGC~7469
\label{table-emspec}}
\tablehead{
\colhead{Line} &\colhead{$\lambda_{\rm vac}$} 
&\colhead{Flux\tablenotemark{1}}
&\colhead{Velocity\tablenotemark{2}}
&\colhead{FWHM} \\
\colhead{} &\colhead{(\AA)} 
&\colhead{}
&\colhead{(\kms)}
&\colhead{(\kms)} }
\startdata
\multicolumn{5}{c}{2002 {\it FUSE}} \\
\hline
O~{\sc vi} broad     &1031.93 &$3.77\pm0.31$ &$-137\pm180$ &$5907\pm380$ \\
O~{\sc vi} broad     &1037.62 &$1.88\pm0.16$ &$-137\pm180$ &$5907\pm380$ \\
O~{\sc vi} narrow    &1031.93 &$2.35\pm0.31$ &$-162\pm54$ &$1061\pm95$ \\
O~{\sc vi} narrow    &1037.62 &$1.17\pm0.16$ &$-162\pm54$ &$1061\pm95$ \\
\hline
\multicolumn{5}{c}{2002 STIS}\\
\hline
Ly$\alpha$ broad    &1215.67 &$17.59\pm0.31$    &$-67\pm89$ &$13170\pm130$ \\
Ly$\alpha$ int.     &1215.67 &$17.13\pm0.49$    &$-202\pm63$      &$4651\pm131$ \\
Ly$\alpha$ narrow1  &1215.67 &$4.89\pm0.25$ &$-319\pm22$      &$1047\pm55$ \\
Ly$\alpha$ narrow2  &1215.67 &$3.26\pm0.52$ &$408\pm50$   &$1586\pm203$ \\
N~{\sc v}   int.     &1240.15 &$5.22\pm0.05$ &$-107\pm33$      &$4651\pm131$ \\
N~{\sc v}   narrow1  &1240.15 &$0.66\pm0.06$ &$-107\pm33$      &$1047\pm55$ \\ 
Si~{\sc ii}          &1263.31 &$0.71\pm0.02$ &$-4\pm93$   &$2723\pm174$ \\
Si~{\sc ii}          &1531.18 &$2.09\pm0.11$ &$-82\pm53$   &$2170\pm88$ \\
C~{\sc iv} broad     &1549.05 &$21.10\pm0.20$    &$224\pm51$   &$9371\pm78$ \\
C~{\sc iv} int.      &1549.05 &$15.57\pm0.06$    &$-105\pm19$      &$3504\pm32$ \\
C~{\sc iv} narrow    &1549.05 &$3.36\pm0.19$ &$-42\pm12$ &$1118\pm47$ \\
Fe~{\sc ii}          &1608.45 &$0.33\pm0.09$ &$795\pm53$   &$2386\pm315$ \\
He~{\sc ii} broad    &1640.50 &$6.86\pm0.45$ &$-326\pm297$     &$11041\pm706$ \\ 
He~{\sc ii} narrow   &1640.50 &$0.89\pm24.41$ &$-326\pm297$ &$1403\pm188$ \\
\hline
\multicolumn{5}{c}{2004 STIS}\\
\hline
Ly$\alpha$ broad    &1215.67 &$4.84\pm0.06$     &$-1315\pm231$  &$13541\pm144$ \\
Ly$\alpha$ int.     &1215.67 &$11.66\pm0.12$    &$-129\pm23$      &$4774\pm54$ \\
Ly$\alpha$ narrow1  &1215.67 &$3.77\pm0.11$ &$-232\pm16$      &$1137\pm19$ \\
Ly$\alpha$ narrow2  &1215.67 &$1.44\pm0.12$ &$449\pm$9   &$1566\pm120$ \\
N~{\sc v}   int.     &1240.15 &$3.84\pm0.09$ &$-38\pm15$      &$4774\pm54$ \\
N~{\sc v}   narrow1  &1240.15 &$0.42\pm0.03$ &$-38\pm15$      &$1137\pm19$ \\
Si~{\sc ii}          &1263.31 &$0.60\pm0.04$ &$-95\pm117$     &$3428\pm276$ \\
Si~{\sc ii}          &1531.18 &$0.95\pm0.04$ &$-82\pm53$   &$2844\pm182$ \\
C~{\sc iv} broad     &1549.05 &$12.13\pm0.27$    &$381\pm77$   &$9069\pm149$ \\
C~{\sc iv} int.      &1549.05 &$8.25\pm0.13$    &$-7\pm32$      &$3657\pm38$ \\
C~{\sc iv} narrow    &1549.05 &$3.97\pm0.12$ &$-58\pm18$ &$1458\pm17$ \\
He~{\sc ii} broad    &1640.50 &$2.79\pm0.75$ &$-326\pm297$     &$10561\pm327$ \\
He~{\sc ii} narrow   &1640.50 &$0.38\pm0.25$ &$-326\pm297$ &$1479\pm913$ \\
\enddata
\tablenotetext{1}{Flux in units of 10$^{-13}$ ergs cm$^{-2}$ s$^{-1}$}
\tablenotetext{2}{Velocity relative to systemic redshift, $z=0.01639$}
\end{deluxetable}

\begin{deluxetable}{lcccccc}
\tablecolumns{5}
\tablewidth{24pc}
\tablecaption{Intrinsic Absorption 
in {\it FUSE} and STIS Spectra of NGC~7469
\label{table-abs}}
\tablehead{
\colhead{Feature} &\colhead{Comp} &\colhead{$\lambda_{\rm vac}$} 
&\colhead{$N_{\rm ion}$}
&\colhead{$\Delta v$\tablenotemark{1}} \\
\colhead{} &\colhead{\# } &\colhead{(\AA)} 
&\colhead{(10$^{12}$ cm$^{-2}$)}
&\colhead{(\kms)}}
\startdata
\multicolumn{5}{c}{2002}\\
\hline
O~{\sc vi}  &2a &1031.93 &$19.2\pm2.0$ &$-2014\pm18$ \\
O~{\sc vi}  &2b &1031.93 &$169\pm17$   &$-1971\pm2$  \\
O~{\sc vi}  &2c &1031.93 &$690\pm71$   &$-1897\pm2$  \\
O~{\sc vi}  &2d &1031.93 &$68.7\pm7.0$ &$-1838\pm18$ \\

O~{\sc vi}  &1a &1031.93 &$160\pm15$   &$-730\pm13$  \\
O~{\sc vi}  &1b &1031.93 &$194\pm18$   &$-628\pm15$  \\
O~{\sc vi}  &1c &1031.93 &$439\pm41$   &$-594\pm9$   \\
O~{\sc vi}  &1d &1031.93 &$753\pm70$   &$-546\pm9$   \\
O~{\sc vi}  &1e &1031.93 &$309\pm29$   &$-490\pm12$  \\
O~{\sc vi}  &1f &1031.93 &$97.9\pm9.1$ &$-483\pm22$  \\
O~{\sc vi}  &1g &1031.93 &$93.4\pm8.7$ &$-426\pm34$  \\
O~{\sc vi}  &1h &1031.93 &$76.8\pm7.1$ &$-397\pm23$  \\
O~{\sc vi}  &1i &1031.93 &$107\pm10$   &$-320\pm13$  \\

Ly$\alpha$ &2a &1215.67 &$5.5\pm3.1$  &$-2005\pm18$ \\
Ly$\alpha$ &2b &1215.67 &$3.8\pm3.1$  &$-1960\pm39$ \\
Ly$\alpha$ &2c &1215.67 &$200\pm58$   &$-1897\pm2$ \\
Ly$\alpha$ &2d &1215.67 &$19.9\pm4.3$ &$-1841\pm5$ \\

Ly$\alpha$ &1a &1215.67 &$23.3\pm3.5$  &$-736\pm4$ \\
Ly$\alpha$ &1b &1215.67 &$28.1\pm12.4$ &$-669\pm15$ \\
Ly$\alpha$ &1c &1215.67 &$63.6\pm31.4$ &$-627\pm14$ \\
Ly$\alpha$ &1d &1215.67 &$109\pm26.6$  &$-567\pm6$ \\
Ly$\alpha$ &1e &1215.67 &$44.8\pm9.8$  &$-504\pm13$ \\
Ly$\alpha$ &1f &1215.67 &$13.5\pm9.1$  &$-460\pm11$ \\
Ly$\alpha$ &1g &1215.67 &$14.3\pm10.4$ &$-422\pm22$ \\
Ly$\alpha$ &1h &1215.67 &$15.5\pm1.9$  &$-367\pm14$ \\
Ly$\alpha$ &1i &1215.67 &$11.1\pm4.3$  &$-316\pm19$ \\

N~{\sc v}   &2a &1238.82  &$6.7\pm0.3$  &$-2014\pm18$ \\
N~{\sc v}   &2b &1238.82  &$2.7\pm3.8$  &$-1971\pm2$ \\
N~{\sc v}   &2c &1238.82  &$242\pm11$   &$-1897\pm2$ \\
N~{\sc v}   &2d &1238.82  &$24.1\pm1.1$ &$-1838\pm18$ \\

N~{\sc v}   &1a &1238.82  &$11.4\pm1.7$  &$-730\pm13$ \\
N~{\sc v}   &1b &1238.82  &$38.0\pm17.0$ &$-618\pm15$ \\
N~{\sc v}   &1c &1238.82  &$18.4\pm18.6$ &$-594\pm9$ \\
N~{\sc v}   &1d &1238.82  &$53.5\pm8.0$  &$-546\pm9$ \\
N~{\sc v}   &1e &1238.82  &$22.0\pm3.3$  &$-490\pm12$ \\
N~{\sc v}   &1f &1238.82  &$6.6\pm1.0$   &$-483\pm22$ \\
N~{\sc v}   &1g &1238.82  &$7.0\pm1.0$   &$-426\pm34$ \\
N~{\sc v}   &1h &1238.82  &$5.4\pm0.8$   &$-397\pm23$ \\
N~{\sc v}   &1i &1238.82  &$7.6\pm1.1$   &$-320\pm13$ \\

C~{\sc iv}  &2a &1548.20  &$5.3\pm0.2$   &$-2014\pm18$ \\
C~{\sc iv}  &2b &1548.20  &$2.0\pm0.1$   &$-1971\pm2$ \\
C~{\sc iv}  &2c &1548.20  &$190\pm8$     &$-1897\pm2$ \\
C~{\sc iv}  &2d &1548.20  &$18.9\pm0.8$  &$-1838\pm18$ \\

C~{\sc iv} &1a &1548.20  &$1.6\pm0.4$   &$-730\pm13$ \\
C~{\sc iv} &1b &1548.20  &$5.5\pm1.4$   &$-618\pm15$ \\
C~{\sc iv} &1c &1548.20  &$2.6\pm0.7$   &$-594\pm9$ \\
C~{\sc iv} &1d &1548.20  &$7.6\pm1.9$   &$-546\pm9$ \\
C~{\sc iv} &1e &1548.20  &$3.1\pm0.8$   &$-490\pm12$ \\
C~{\sc iv} &1f &1548.20  &$0.9\pm0.2$   &$-483\pm22$ \\
C~{\sc iv} &1g &1548.20  &$1.0\pm0.2$   &$-426\pm34$ \\
C~{\sc iv} &1h &1548.20  &$0.8\pm0.2$   &$-397\pm23$ \\
C~{\sc iv} &1i &1548.20  &$1.1\pm0.3$   &$-320\pm13$ \\

\hline
\multicolumn{5}{c}{2004}\\
\hline
Ly$\alpha$ &2a &1215.67  &$6.4\pm3.0$  &$-2005\pm18$ \\
Ly$\alpha$ &2b &1215.67  &$5.1\pm3.1$  &$-1959\pm38$ \\
Ly$\alpha$ &2c &1215.67  &$191\pm36$   &$-1897\pm2$ \\
Ly$\alpha$ &2d &1215.67  &$36.3\pm5.4$ &$-1841\pm5$ \\

Ly$\alpha$ &1a &1215.67  &$58.1\pm6.1$ &$-736\pm4$ \\
Ly$\alpha$ &1b &1215.67  &$0.09\pm5$   &$-669\pm15$ \\
Ly$\alpha$ &1c &1215.67  &$304\pm89$   &$-627\pm14$ \\
Ly$\alpha$ &1d &1215.67  &$226\pm36$   &$-567\pm6$ \\
Ly$\alpha$ &1e &1215.67  &$83.7\pm10.3$ &$-504\pm13$ \\
Ly$\alpha$ &1f &1215.67  &$9.8\pm1.4$  &$-460\pm11$ \\
Ly$\alpha$ &1g &1215.67  &$31.7\pm4.0$ &$-422\pm22$ \\
Ly$\alpha$ &1h &1215.67  &$14.9\pm3.2$ &$-367\pm14$ \\
Ly$\alpha$ &2i &1215.67  &$7.9\pm2.6$  &$-316\pm19$ \\

N~{\sc v}   &2a &1238.82  &$6.0\pm0.4$  &$-2014\pm18$ \\
N~{\sc v}   &2b &1238.82  &$8.9\pm5.9$  &$-1971\pm2$ \\
N~{\sc v}   &2c &1238.82  &$216\pm15$   &$-1897\pm2$ \\
N~{\sc v}   &2d &1238.82  &$21.5\pm1.5$ &$-1838\pm18$ \\

N~{\sc v}   &1a &1238.82  &$19.0\pm3.0$  &$-730\pm13$ \\
N~{\sc v}   &1b &1238.82  &$64.0\pm23.6$ &$-618\pm15$ \\
N~{\sc v}   &1c &1238.82  &$41.0\pm26.0$ &$-594\pm9$ \\
N~{\sc v}   &1d &1238.82  &$89.0\pm14.1$ &$-546\pm9$ \\
N~{\sc v}   &1e &1238.82  &$36.5\pm5.8$  &$-490\pm12$ \\
N~{\sc v}   &1f &1238.82  &$11.0\pm1.7$  &$-483\pm34$ \\
N~{\sc v}   &1g &1238.82  &$11.6\pm1.8$  &$-426\pm22$ \\
N~{\sc v}   &1h &1238.82  &$9.1\pm1.4$   &$-399\pm13$ \\
N~{\sc v}   &1i &1238.82  &$12.6\pm2.0$  &$-320\pm23$ \\

C~{\sc iv} &2a &1548.20  &$4.8\pm0.3$   &$-2014\pm18$ \\
C~{\sc iv} &2b &1548.20  &$1.8\pm0.1$   &$-1971\pm2$ \\
C~{\sc iv} &2c &1548.20  &$173\pm12$    &$-1897\pm2$ \\
C~{\sc iv} &2d &1548.20  &$17.2\pm1.1$  &$-1838\pm18$ \\

C~{\sc iv} &1a &1548.20  &$6.6\pm0.6$  &$-730\pm13$ \\
C~{\sc iv} &1b &1548.20  &$22.2\pm1.9$ &$-618\pm15$ \\
C~{\sc iv} &1c &1548.20  &$10.8\pm0.9$ &$-594\pm9$ \\
C~{\sc iv} &1d &1548.20  &$30.9\pm2.7$ &$-546\pm9$ \\
C~{\sc iv} &1e &1548.20  &$12.7\pm1.1$ &$-490\pm12$ \\
C~{\sc iv} &1f &1548.20  &$3.8\pm0.3$  &$-483\pm22$ \\
C~{\sc iv} &1g &1548.20  &$4.0\pm0.3$  &$-426\pm34$ \\
C~{\sc iv} &1h &1548.20  &$3.1\pm0.3$  &$-397\pm23$ \\
C~{\sc iv} &1i &1548.20  &$4.3\pm0.4$  &$-320\pm13$ \\
\enddata
\tablenotetext{1}{Velocity with respect to
systemic redshift of \ngc, $z=0.01639$ (de Vaucouleurs et al.\ 1991)}
\end{deluxetable}

\clearpage

\clearpage

\clearpage

\clearpage

\clearpage

\clearpage

\clearpage

\clearpage

\clearpage

\clearpage

\clearpage

\clearpage

\clearpage

\clearpage

\clearpage

\clearpage

\clearpage

\clearpage

\clearpage


\begin{thebibliography}

\bibitem[Adelberger, Steidel, Shapley, \& 
Pettini(2003)]{2003ApJ...584...45A} Adelberger, K.~L., Steidel, C.~C., 
Shapley, A.~E., \& Pettini, M.\ 2003, \apj, 584, 45 

\bibitem[Arav et al.(2002)]{2002ApJ...566..699A} Arav, N., Korista, K.~T., 
\& de Kool, M.\ 2002, \apj, 566, 699 

\bibitem[Ballantyne \& Fabian(2001)]{2001MNRAS.328L..11B} Ballantyne, 
D.~R., \& Fabian, A.~C.\ 2001, \mnras, 328, L11

\bibitem[Bechtold et al.(2002)]{2002ApJS..140..143B} Bechtold, J., 
Dobrzycki, A., Wilden, B., Morita, M., Scott, J., Dobrzycka, D., Tran, K., 
\& Aldcroft, T.~L.\ 2002, \apjs, 140, 143 

\bibitem[Behar \& Netzer(2002)]{2002ApJ...570..165B} Behar, E.~\& Netzer, 
H.\ 2002, \apj, 570, 165 

\bibitem[Bianchi \& Matt(2002)]{2002A&A...387...76B} Bianchi, S., \& Matt, 
G.\ 2002, \aap, 387, 76

\bibitem[B]{blustin02} Blustin, A. J., Branduardi-Raymont, G., Behar, E., Kaastra, J. S.,
Kahn, S. M., Page, M. J., Sako, M., \& Steenbrugge, K. C. 2002, \aap, 392, 453

\bibitem[B]{blustin03} Blustin, A. J., et al. 2003, \aap, 403, 481

\bibitem[B]{blustin04} Blustin, A. J., Page, M. J., Fuerst, S. V., Branduardi-Raymont, 
G., \& Ashton, C. E. 2005, \aap, 431, 111

\bibitem[B]{brotherton} Brotherton, M. S., Green, R. F., Kriss, G. A., Oegerle, W., Kaiser,
M. E., Zheng, W., \& Hutchings, J. B. 2002, \apj, 565, 800

\bibitem[CCM]{ccm89} Cardelli, J. A., Clayton, G. C., \& Mathis, J. S. 1989, \apj, 345, 245 

\bibitem[Cavaliere, Lapi, \& Menci(2002)]{2002ApJ...581L...1C} Cavaliere, 
A., Lapi, A., \& Menci, N.\ 2002, \apjl, 581, L1 

\bibitem[Collier et al.(1998)]{1998ApJ...500..162C} Collier, S.~J., et al.\ 
1998, \apj, 500, 162 

\bibitem[Collinge et al.(2001)]{2001ApJ...557....2C} Collinge, M.~J., et 
al.\ 2001, \apj, 557, 2 

\bibitem[C]{ck1999} Crenshaw, D. M. \& Kraemer, S. B. 1999, \apj, 521, 572

\bibitem[Crenshaw et al.\ 1999]{cren1999}
 Crenshaw, D. M., Kraemer, S. B., Boggess, A., Maran, S. P., Mushotzky,
 R. F., \& Wu, C. 1999, \apj, 516, 750

\bibitem[Crenshaw et al.\ 2000]{cren2000}
 Crenshaw, D. M., Kraemer, S. B., Hutchings, J. B.,
 Danks, A. C., Gull, T. R., Kaiser, M. E.,
 Nelson, C. H., \& Weistrop, D. 2000, \apj, 545, L27

\bibitem[Crenshaw et al.(2003)]{2003ApJ...594..116C} Crenshaw, D.~M., et 
al.\ 2003, \apj, 594, 116 

\bibitem[Crenshaw, Kraemer, \& George(2003)]{2003ARA&A..41..117C} Crenshaw, 
D.~M., Kraemer, S.~B., \& George, I.~M.\ 2003, \araa, 41, 117 

\bibitem[De Rosa, Fabian, \& Piro(2002)]{2002MNRAS.334L..21D} De Rosa, A., 
Fabian, A.~C., \& Piro, L.\ 2002, \mnras, 334, L21 

\bibitem[d91]{rc3} de Vaucouleurs, G., de Vaucouleurs, A., Corwin, H. G.,
 Buta, R. J., Paturel, G., \& Fouqu\`{e}, P., 1991, Third Reference Catalog of Bright
 Galaxies, Version 3.9 (New York: Springer Verlag)

\bibitem[dl90]{dl90} Dickey, J. M. \& Lockman F. J. 1990, \araa, 28, 215

\bibitem[Gabel et al.\ 2005]{gabel2005} Gabel, J. R. et al.\ 2005, \apj,
{\it in press}
 
\bibitem[Gabel et al.\ 2003a]{gabel2003a} Gabel, J. R. et al.\ 2003a, \apj, 595, 120

\bibitem[Gabel et al.\ 2003b]{gabel2003b} Gabel, J. R. et al.\ 2003b, \apj, 583, 178

\bibitem[G86]{g86} Gehrels, N. 1986, \apj, 303, 336

\bibitem[G98]{george98} George, I. M., Turner, T. J., Netzer, H.,
 Nandra, K., Mushotzky, R. F., \& Yaqoob, T. 1998, \apjs, 114, 73

\bibitem[Granato et al.(2004)]{2004ApJ...600..580G} Granato, G.~L., De 
Zotti, G., Silva, L., Bressan, A., \& Danese, L.\ 2004, \apj, 600, 580 

\bibitem[Grevesse et al.(1996)]{1996coab.proc..117G} Grevesse, N., Noels, 
A., \& Sauval, A.~J.\ 1996, ASP Conf.~Ser.~ 99: Cosmic Abundances, 117 

\bibitem[HF99]{hf99} Hamann, F. \& Ferland, G. 1999, \araa, 37, 487

\bibitem[Houck]{} 
Houck, J.~C. \& Denicola, L.~A. 2000, in ASP Conf. Ser. 216: Astronomical
  Data Analysis Software and Systems IX, 591

\bibitem[Kaastra et al.(2000)]{2000A&A...354L..83K} Kaastra, J.~S., Mewe, 
R., Liedahl, D.~A., Komossa, S., \& Brinkman, A.~C.\ 2000, \aap, 354, L83 

\bibitem[K02]{k02} Kaastra, J. S., Steenbrugge, K. C., Raassen, A. J. J.,
 van der Meer, R. L. J.,
 Brinkman, A. C., Leidahl, D. A., Behar, E., \& de Rosa, A. 2002, \aap, 427, 2002

\bibitem[K00]{k00}
Kallman, T. R. 2000, XSTAR User's Guide (Greenbelt, MD: NASA/GSFC) 

\bibitem[K01]{k01} Kaspi, S., et al. 2001, \apj, 554, 216

\bibitem[K2001]{kraemer2001} Kraemer, S. B., Crenshaw, D. M., \& Gabel, J. R. 2001, \apj, 557, 30

\bibitem[K2001a]{kraemer2001a} Kraemer, S. B. et al. 2001, \apj, 551, 671

\bibitem[K2002a]{kraemer2002} Kraemer, S. B., Crenshaw, D. M., George, I. M., Netzer, H.,
 Turner, T. J., \& Gabel, J. R. 2002, \apj, 577, 98

\bibitem[K2003]{kraemer2003} Kraemer, S. B., Crenshaw, D. M., Yaqoob, T.,
  M$^{\rm c}$Kernan, B.,  Gabel, J. R.,  George, I. M.,  Turner, T. J., \& Dunn, J. P. 2003,
  \apj, 582, 125

\bibitem[]{KK95} Krolik, J.~H. \& Kriss, G.~A. 1995, \apj, 447, 512

\bibitem[]{KK02} Krolik, J.~H. \& Kriss, G.~A. 2001, \apj, 561, 684

\bibitem[K1994]{kriss1994} Kriss, G. A. 1994, in ASP Conf. Ser. 61, Astronomical
 Data Analysis Software and Systems III, ed. Dr. R. Crabtree, R. J. Hanisch, \&
 J. Barnes (San Francisco: ASP), 437 

\bibitem[K1995]{kriss1995} Kriss, G.~A., Davidsen, A.~F., Zheng, W., Kruk, J.~W., \& Espey,
B.~R. 1995, \apj, 454, L7

\bibitem[K2003]{kriss2003} Kriss, G. A., Blustin, A., Branduardi-Raymont, G., Green,
R. F., Hutchings, J., \& Kaiser, M. E. 2003, \aap, 403, 473

\bibitem[K2000a]{kriss2000a} Kriss, G. A., Peterson, B. M., Crenshaw, D. M., \& Zheng, W.
 2000a, \apj, 535, 58

\bibitem[K2000b]{kriss2000b} Kriss, G. A. et al. 2000b, \apj, 538, L17

\bibitem[Krongold et al.(2005)]{2005ApJ...622..842K} Krongold, Y., 
Nicastro, F., Brickhouse, N.~S., Elvis, M., \& Mathur, S.\ 2005, \apj, 622, 
842 

\bibitem[Lee et al.(1999)]{1999MNRAS.310..973L} Lee, J.~C., Fabian, A.~C., 
Brandt, W.~N., Reynolds, C.~S., \& Iwasawa, K.\ 1999, \mnras, 310, 973 

\bibitem[Leitherer et al.(2001)]{2001ApJ...550..724L} Leitherer, C., Le{\~ 
a}o, J.~R.~S., Heckman, T.~M., Lennon, D.~J., Pettini, M., \& Robert, C.\ 
2001, \apj, 550, 724 

\bibitem[Makishima et~al.1986]{xspecdiskbb2}
Makishima, K., Maejima, Y., Mitsuda, K., Bradt, H.~V., Remillard,
  R.~A., Tuohy, I.~R., Hoshi, R., \& Nakagawa, M. 1986, \apj, 308, 635

\bibitem[Marsh2]{marsh2} Marshall H.~L., Dewey, D., \& Ishibashi, K. 2004a,
SPIE, 5165, 457

\bibitem[Marsh]{marsh} Marshall H.~L., Tennant A., Grant C.~E., Hitchcock A.~P., O'Dell S.,
Plucinsky P.~P. 2004b, SPIE, 5165, 497

\bibitem[M1995]{mathur1999}Mathur, S., Elvis, M., \& Wilkes, B. 1999, \apj, 519, 605

\bibitem[Mitsuda et~al.]{xspecdiskbb1}
Mitsuda, K., et~al. 1984, \pasj, 36, 741

\bibitem[Nandra et al.(1998)]{1998ApJ...505..594N} Nandra, K., Clavel, J., 
Edelson, R.~A., George, I.~M., Malkan, M.~A., Mushotzky, R.~F., Peterson, 
B.~M., \& Turner, T.~J.\ 1998, \apj, 505, 594 

\bibitem[Nandra et al.(2000)]{2000ApJ...544..734N} Nandra, K., Le, T., 
George, I.~M., Edelson, R.~A., Mushotzky, R.~F., Peterson, B.~M., \& 
Turner, T.~J.\ 2000, \apj, 544, 734 

\bibitem[Netzer et al.(2003)]{2003ApJ...599..933N} Netzer, H., et al.\ 
2003, \apj, 599, 933 

\bibitem[Ogle et al.(2004)]{2004ApJ...606..151O} Ogle, P.~M., Mason, K.~O., 
Page, M.~J., Salvi, N.~J., Cordova, F.~A., McHardy, I.~M., \& Priedhorsky, 
W.~C.\ 2004, \apj, 606, 151 

\bibitem[reynolds]{r97} Reynolds, C. S. 1997, MNRAS, 286, 513

\bibitem[Sahnow]{sahnow2000} Sahnow, D. J., et al. 2000, ApJ, 538, L7 

\bibitem[Scannapieco \& Oh(2004)]{2004ApJ...608...62S} Scannapieco, E.~\& 
Oh, S.~P.\ 2004, \apj, 608, 62 

\bibitem[Schlegel, Finkbeiner, \& Davis 1998]{schlegel} Schlegel, D. J., Finkbeiner,
 D. P., \& Davis, M. 1998, \apj, 500, 525

\bibitem[Schurch et al.(2003)]{2003MNRAS.345..423S} Schurch, N.~J., 
Warwick, R.~S., Griffiths, R.~E., \& Sembay, S.\ 2003, \mnras, 345, 423

\bibitem[Scott et al.(2004)]{2004ApJS..152....1S} Scott, J.~E., et al.\ 
2004, \apjs, 152, 1 

\bibitem[Steenbrugge et al. 2003]{s03} Steenbrugge, K. C., Kaastra, J. S.,
 de Vries, C. P., \& Edelson, R. 2003, A\&A, 402, 477

\bibitem[Wakker et al.(2003)]{2003ApJS..146....1W} Wakker, B.~P.~et al.\ 
2003, \apjs, 146, 1 

\bibitem[W]{wakker2001} Wakker, B. P., Kalberla, P. M. W., van Woerden, H.,
 de Boer, K. S., \& Putman, M. E. 2001, \apjs, 136, 537

\bibitem[Wandel, Peterson, \& Malkan(1999)]{1999ApJ...526..579W} Wandel, 
A., Peterson, B.~M., \& Malkan, M.~A.\ 1999, \apj, 526, 579 

\bibitem[Wanders et al.(1997)]{1997ApJS..113...69W} Wanders, I., et al.\ 
1997, \apjs, 113, 69 

\bibitem[Wilms et~al. 2000]{ }
Wilms, J., Allen, A., \& M${\rm c}$Cray, R. 2000, \apj, 542, 914

\bibitem[Y]{yaqoob} Yaqoob, T., M$^{\rm c}$Kernan, B., Kraemer, S. B., Crenshaw, D. M.,
Gabel, J. R., George, I. M., \& Turner, T. J. 2003, \apj,  582, 105

\end{thebibliography}
\end{document}